\documentclass[final,1p,times,twocolumn]{elsarticle}

\usepackage{amssymb}
\usepackage{graphicx}
\usepackage{subfig}

\def\MLine#1{\par\hspace*{-\leftmargin}\parbox{\textwidth}{\[#1\]}}

\newcommand{\Proteus}{{\sc Proteus}}
\newcommand{\FLASH}{{\sc FLASH}}
\newcommand{\etal}{{et al.\/}\/ }

\journal{High Energy Density Physics}

\begin{document}

\begin{frontmatter}



\title{The Braginskii model of the Rayleigh-Taylor
  instability. I. Effects of self-generated magnetic fields and
  thermal conduction in two dimensions}

%
%
%
\author[fsu]{Frank Modica}

\author[fsu]{Tomasz Plewa\corref{cor1}}
\ead{tplewa@fsu.edu}

\author[kit]{Andrey Zhiglo}
\ead{azhiglo@gmail.com}

\cortext[cor1]{Corresponding Author: Florida State University,
  Department of Scientific Computing, 415 Dirac Science Library,
  Tallahassee, FL 32306-4120, USA.  Phone: (850) 644-1959 Fax: (850)
  644-0098, Email: tplewa@fsu.edu}

\address[fsu]{Department of Scientific Computing\\ Florida State University, Tallahassee, FL, USA}

\address[kit]{NSC Kharkov Institute of Physics and Technology\\ Kharkov, Ukraine}
\begin{abstract}

There exists a substantial disagreement between computer simulation
results and high-energy density laboratory experiments of the
Rayleigh-Taylor instability \cite{kuranz+10}. Motivated by the
observed discrepancies in morphology and growth rates, we attempt to
bring simulations and experiments into better agreement by extending
the classic purely hydrodynamic model to include self-generation of
magnetic fields and anisotropic thermal conduction.

We adopt the Braginskii formulation for transport in hot, dense
plasma, implement and verify the additional physics modules, and
conduct a computational study of a single-mode RTI in two dimensions
with various combinations of the newly implemented modules. We analyze
physics effects on the RTI mixing and flow morphology, the effects of
mutual physics interactions, and the evolution of magnetic fields.

We find that magnetic fields reach levels on the order of $11$ MG
(plasma $\beta\approx{9.1\times{10^{-2}}}$) in the absence of thermal
conduction. These fields do not affect the growth of the mixed layer
but substantially modify its internal structure on smaller scales. In
particular, we observe denting of the RT spike tip and generation of
additional higher order modes as a result of these fields. Contrary to
interpretation presented in earlier work \cite{nishiguchi02}, the
additional mode is not generated due to modified anisotropic heat
transport effects but due to dynamical effect of self-generated
magnetic fields. The overall flow morphology in self-magnetized,
non-conducting models is qualitatively different from models with a
pre-existing uniform field oriented perpendicular to the
interface. This puts the usefulness of simple MHD models for
interpreting the evolution of self-magnetizing HED systems with
zero-field initial conditions into doubt.

The main effects of thermal conduction are a reduction of the RT
instability growth rate (by about 20\% for conditions considered here)
and inhibited mixing on small scales. In this case, the maximum
self-generated magnetic fields are weaker (approximately ${1.7}$ MG;
plasma $\beta\approx{49}$). This is due to reduction of temperature
and density gradients due to conduction. These self-generated magnetic
fields are of very similar strength compared to magnetic fields
observed recently in HED laboratory experiments \cite{manuel+12}.

We find that thermal conduction plays the dominant role in the
evolution of the model RTI system considered. It smears out
small-scale structure and reduces the RTI growth rate. This may
account for the relatively featureless RT spikes seen in experiments,
but does not explain mass extensions observed in experiments.

Resistivity and related heat source terms were not included in the
present work, but we estimate their impact on RTI as modest and not
affecting our main conclusions. Resistive effects will be discussed in
detail in the next paper in the series.

\end{abstract}
\begin{keyword}
magnetohydrodynamics \sep hydrodynamic instabilities \sep thermal conduction \sep  Braginskii equations \sep laboratory astrophysics
\end{keyword}

\end{frontmatter}
%
%
%
\section{Introduction}\label{s:intro}
The Rayleigh-Taylor instability (RTI) occurs when a dense fluid is
accelerated by a light fluid. At early times, RTI produces
characteristic morphology with heavy material penetrating into a light
fluid in forms of fingers or spikes which are well-separated by
bubbles filled with the light fluid \cite{strutt83,taylor50}. The
resulting material mixing plays an important and sometimes critical
role in many problems including industrial applications
\cite{debacq+01,lawrie+11}, inertial confinement fusion
\cite{lindl+04,clark+11}, and evolution of stellar objects such as
supernovae and their remnants (see
\cite{arnett+89,khokhlov95,blondin+01} and references therein). In the
case of core-collapse supernovae (ccSNe), RTI occurs when a supernova
shock expands with a varying speed through the progenitor envelope
(see, e.g., \cite{gawryszczak+10} and references therein).

Conditions relevant to mixing of the chemical elements in ccSNe were
successfully reproduced in a series of high-energy density (HED)
laboratory experiments \cite{remington+97,drake+02,kuranz+10}. In the
HED experiments, the RT instability is created when a laser-driven
blast wave moves through either planar or spherical target from a high
to low-density material. The target is designed to mimic a structure
characteristic of compositional interfaces in a ccSN progenitor
envelope. Experiments are routinely aided by extensive computational
studies at both the design and analysis stages.

In this paper, we focus on unusual morphological features of RTI
observed in the \citet{kuranz+10} experiment. The reported RTI
morphology was significantly different from the results of
hydrodynamic simulations, showing strongly suppressed growth of small
scale structure and mass extensions of RT spikes. In their analysis,
\citet{kuranz+10} focused on the discussion of RT spike mass
extensions. They concluded that pure hydrodynamic models cannot
account for the observed features and theorized that magnetic fields
might be responsible for extended structures emanating from tips of RT
spikes.

Large-scale magnetic fields are known to affect the RTI dynamics and
its growth rate \cite{chandrasekhar61}. Simulations performed by
\citet{jun+95} demonstrated that magnetic fields reduced RTI growth
for single-mode perturbations, in good agreement with linear
theory. However, in the case of multi-mode perturbations they observed
faster growth of RTI when the field was tangential to the
interface. They also found stronger amplification of magnetic fields
in 3-D than in 2-D. More recently, \citet{stone+07} found that in 3-D,
strong magnetic fields parallel to the interface increase RTI growth
rates by inhibiting sheer at the interface and suppressing material
mixing. Given that neither of the above studies indicated presence of
RT spike mass extensions, \citet{kuranz+10} proposed that the target
is self-magnetized. The relevant physical process is known as the
Biermann battery effect \cite{biermann50,kulsrud05}. In it, a
thermoelectric current is spontaneously created whenever the electron
temperature and electron density gradients are misaligned and results
in \mbox{(self-)}generation of magnetic field. This process does not require
any seed magnetic field to operate. The in situ generated field can
subsequently be modified due to fluid flow and additional interactions
between ions and electrons in the plasma.

Thermal conduction is also known to have varying effects on plasma
dynamics.  Numerous studies have shown that heat transfer can decrease
RTI growth rates \cite{Bychkov1994,Betti1998}. The opposite effect was
found by \citet{Ryutov2000} who demonstrated that it is possible to
create conditions under which thermal conduction actually has a
destabilizing effect and induces RTI. When plasma is magnetized, it is
expected that heat transfer by electrons across magnetic field lines
becomes suppressed if the electron gyroradius gets smaller than its
mean free path \cite{braginskii65}. \citet{Lecoanet2012} studied
systems stable and unstable against RTI. They showed that anisotropic
heat transport affects only compressible modes, and these modes can
actually grow faster.

In the context of inertial confinement fusion (ICF),
\citet{nishiguchi02} found the magnetic fields generated by RTI strong
enough to make the electron thermal conduction anisotropic. In
particular, Nishiguchi found thermal conduction to be most efficient
near the center of a RT spike. Nishiguchi also noted that an
additional mode was growing at the interface and claimed conduction
responsible for this new flow feature. Generated magnetic fields in
megagauss range were obtained in simulations by several groups
\citet{nishiguchi02,mima+78}; comparable field level was observed
recently in experiments \citet{manuel+12,gao+12}.

In this work, motivated by the recent results of laser-driven
high-energy density experiments of \citet{kuranz+10}, we study the
effects of self-generated magnetic fields and thermal conduction on
the Rayleigh-Taylor instability in a basic plasma physics setting in
two spatial dimensions. Our physics model is a simplified version of
the Braginskii model of a single fluid MHD. We implement and verify
suitable additional physics modules in our hydrocode and perform a
series of simulations to assess numerical model convergence. In
application to RTI, we consider conditions relevant to the Kuranz
\etal experiment, and focus on individual physics effects and their
interactions. We discuss the results and propose directions for the
future research in this area.
\section{Computational methods}\label{s:meth}
\subsection{Extended MHD model with anisotropic conduction}\label{sec:ExtMHD}
An electrically conducting fluid can be described by one-fluid and
two-fluid plasma models. In a one-fluid plasma model, fluid flow
equations are written in terms of the average fluid velocity only, and
are coupled with Maxwell's equations of electromagnetism via Ohm's
law. This description is adequate when the plasma can be considered
locally electrically neutral. In a two-fluid plasma model, the
dynamics of electrons and ions are described individually, for
example, by the Braginskii equations \cite{braginskii65}. This is a
more accurate description of the plasma than a one-fluid model, but is
also more expensive requiring to solve twice as many fluid flow
equations.

In this paper we numerically solve the magnetohydrodynamic (MHD)
equations in conservation form using the \Proteus\ code. \Proteus\ is
based on the \FLASH\ code \cite{fryxell+00}, and includes additional
physics and algorithms developed to model anisotropic thermal
conduction and magnetic field generation. This extended physics model
is based on a single-fluid formulation by \citet{braginskii65}, and
described by the following set of equations,
\begin{eqnarray*}
\!\!\!\!\!\!\!&\!\!\!\!\!& \!\!\!\!\!\!\!\!\!     \frac{\partial\rho}{\partial t} + \nabla\cdot\rho\vec{V} = 0,\\
\!\!\!\!\!\!\!&\!\!\!\!\!& \!\!\!\!\!\!\!\!\!     \nabla \cdot \vec{B} = 0,\\
\!\!\!\!\!\!\!&\!\!\!\!\!& \!\!\!\!\!\!\!\!\!     \frac{\partial\rho\vec{V}}{\partial t} +
        \nabla\cdot\left(\rho\vec{V}\cdot\vec{V}
        -\frac{\vec{B}\cdot\vec{B}}{4\pi}\right) + \nabla P = \rho \vec{g},\\
\!\!\!\!\!\!\!&\!\!\!\!\!& \!\!\!\!\!\!\!\!\!     \frac{\partial\rho E}{\partial t} + \nabla\cdot\left( (\rho E
        + P) \vec{V} -\frac{\vec{B}\vec{B}}{4\pi} \cdot \vec{V}
        \right) = \rho\vec{V}\cdot \vec{g} + \nabla\cdot({\vec{q}_T^{\,e}}
        +{\vec{q}_T^{\,i}}),\\
\!\!\!\!\!\!\!&\!\!\!\!\!& \!\!\!\!\!\!\!\!\!     \frac{\partial\vec{B}}{\partial t} =
        \nabla\times(\vec{V}\times\vec{B}) +
        \frac{c}{e}\left[\nabla\times\frac{\nabla P_e}{n_e}  -
          \nabla\times\frac{(\nabla\times\vec{B})\times\vec{B}}{4 \pi
            n_e}\right. \\
 &\!\!\!&\quad\!\!\!\!\!\!       \left. -\: \nabla{\times}\frac{\vec{R}_T}{n_e}  \right],
\end{eqnarray*}
In these equations, $\rho$ is the plasma density, $\vec{V}$ is the
plasma velocity, $E$ is the total plasma specific energy, $P$ is the plasma
pressure, and $\vec{B}$ is the magnetic field. We also include effects
due to constant gravitational acceleration, $\vec{g}$. In the last
equation above, $n_e$ is the electron number density, $P_e$ is the
electron pressure, and other symbols have their usual meaning. The
last term inside the square brackets above is simplified compared to the
Braginskii model, and only includes a contribution due to the thermal
force,
\[
\vec{R}_T = -\beta_\|^{uT} \nabla_\|T_e -\beta_\perp^{uT} \nabla_\perp T_e -\beta_\wedge^{uT}[\vec{h} {\times}\nabla T_e].
\]
Here $T_e$ is the electron temperature, $\nabla_\|$ and $\nabla_\perp$
are mean components of the gradient operator (here acting on the
electron temperature) parallel and perpendicular to the magnetic
field, and $\vec{h}$ is a unit vector parallel to the magnetic field.

The last term on the right hand side of the energy equation describes
thermal conduction, where $\vec{q}_T^{\,e}$ and $\vec{q}_T^{\,i}$ are the
electron and ion heat flux vectors, respectively:
\begin{eqnarray*}
{\vec{q}_T^{\,e}} &=& -\kappa_\|^e \nabla_\|T_e -\kappa_\perp^e \nabla_\perp T_e -\kappa_\wedge^{e}[\vec{h} {\times}\nabla T_e],\\
{\vec{q}_T^{\,i}} &=& -\kappa_\|^i \nabla_\|T_i -\kappa_\perp^i \nabla_\perp T_i +\kappa_\wedge^{i}[\vec{h} {\times}\nabla T_i],
\end{eqnarray*}
where $T_i$ is the temperature of ions.  In agreement with the
Braginskii model, we assume both species are in thermal equilibrium,
$T_e = T_i$. The electron heat conduction coefficients are
\begin{equation}
\begin{array}{lcl}
\displaystyle \kappa_\|^e &=& \frac{n_e T_e \tau_e}{m_e} \gamma_0,\\
\displaystyle \kappa_\perp^e &=& \frac{n_e T_e \tau_e}{m_e} \frac{(\gamma_1^{\prime}\chi^2+\gamma_0^{\prime})}{\Delta_e},\\
\displaystyle \kappa_\wedge^e &=& \frac{n_e T_e \tau_e}{m_e} \frac{\chi(\gamma_1^{\prime\prime}\chi^2+\gamma_0^{\prime\prime})}{\Delta_e},
\end{array}
\label{e:kappa_e}
\end{equation}
and the ion conduction coefficients are
\begin{equation}
\begin{array}{lcl}
\displaystyle \kappa_\|^i &=& \frac{3.906 n_i T_i \tau_i}{m_i},\\
\displaystyle \kappa_\perp^i &=& \frac{n_i T_i \tau_i}{m_i} \frac{(2\chi^2 + 2.645)}{\Delta_i},\\
\displaystyle \kappa_\wedge^i &=& \frac{n_i T_i \tau_i}{m_i} \frac{\chi(\frac{5}{2}\chi^2 + 4.65)}{\Delta_i}.
\end{array}
\label{e:kappa_i}
\end{equation}
Here $\Delta_e = \chi_e^4 + \delta_1 \chi_e^2 +\delta_0$ and $\Delta_i
= \chi_i^4 + 2.70 \chi_i^2 +0.677$, with $\chi_e = \omega_e \tau_e$
and $\chi_i = \omega_i \tau_i$. $\omega_e$ and $\omega_i$ are the
cyclotron frequency of electrons and ions, respectively,
\begin{eqnarray*}
\omega_e &=& \frac{e B}{m_e c},\\
\omega_i &=& \bar Z \frac{e B}{m_i c},
\end{eqnarray*}
and $\tau_e$ and $\tau_i$ are the ion-electron and ion-ion collision times, respectively,
\begin{eqnarray*}
\tau_e &=& \frac{3 {\sqrt{m_e}} {T_e}^{\frac{3}{2}} }{4 {\sqrt{2\pi}}\lambda e^4 {\bar{Z}^2 n_i}},\\
\tau_i &=& \frac{3 {\sqrt{m_i}} {T_i}^{\frac{3}{2}} }{4 {\sqrt{\pi}}\lambda e^4 {\bar{Z}^4 n_i}}.
\end{eqnarray*}
$\lambda$ is the Coulomb logarithm; all other symbols are defined in \cite{braginskii65}.

Compared to the original Braginskii formulation, our model does not
take into account electric resistivity effects. This is primarily due
to severe time step restriction imposed by corresponding friction
force term in the induction equations, which would make computations
infeasible. However, our results and estimates demonstrate resistivity
will not affect the dynamics of the system considered here or change
our main conclusions. Also, \citet{budde+10} found that viscosity
plays only a minor role in the evolution of RTI targets in the HED
experiments. Consequently, we decided not to included viscous effects
in this work. Omission of the above two physics processes is the only
difference between our model and the Braginskii formulation.

We used the ideal equation of state with $\gamma = 5/3$, and assumed a
plasma composed of a single species with atomic mass $A$, and atomic
charge $Z$. In this case, the ion number density is $n_i = \rho
N_\mathrm{A}/A$, $N_\mathrm{A}$ being the Avogadro constant. In order
to calculate the electron number density $n_e$, we first compute the
(average) plasma charge, $\bar Z$, using the Thomas-Fermi equation of
state \cite{Salzmann1998}. The corresponding electron number density
is $n_e={\bar Z}n_i$.
\subsection{Implementation of the model}

The additional source terms for self-generation of magnetic fields and
anisotropic thermal conduction were implemented in \Proteus\ and
integrated with the directionally unsplit MHD \FLASH\ solver, USM
\cite{lee+09}. The required modifications to the USM algorithm are
rather modest. Assuming 2-D planar geometry, we proceed as follows:
\begin{itemize}

\item At the beginning of the time step, $n$, compute a stable time
  step, $\Delta t$, based on thermal conductivity coefficients and
  magnetic field growth rates.

\item Apply source terms to magnetic fields at cell centers:

\MLine{\vec{B}_{i,j}^{n+\frac{1}{2}} = \vec{B}_{i,j}^{n} + \frac{\Delta t}{2}\frac{\partial\vec{B}_{i,j}^{n}}{\partial t}.}

\item Compute fluxes at cell interfaces at half time step using the USM solver.

\item Compute heat fluxes due to thermal conduction at cell interfaces and add them to the total energy fluxes.

\item Execute the remaining part of the USM solver to advance the
  solution to the next time level, $t^{n+1} = t^{n}$.

\item Using the updated state, compute the magnetic field source terms
  at cell centers and correct the magnetic fields:

\MLine{\vec{B}_{i,j}^{n+1}  \rightarrow \vec{B}_{i,j}^{n+1} + \frac{\Delta t}{2}\frac{\partial\vec{B}_{i,j}^{n+1}}{\partial t}.}

\item End of step; advance the time step counter, $n \rightarrow n + 1$.

\end{itemize}
The extension of the above algorithm to 3-D is straightforward. The
algorithm is explicit and first-order accurate in time. A higher-order
method would be much desired, but considerably more complex to
implement and its development is beyond the scope of the current
paper.
\subsection{Solution verification tests}
We performed several test in order to verify that our extended code
correctly generates magnetic fields and accounts for the effects
of anisotropic heat conduction. To this end, we designed a simple test
to verify that differential operators produce correct results and used
problems with known exact solutions to assess performance of the
Biermann battery source term and anisotropic thermal conduction
solver. In the following sections we present the results of the above
solution verification tests. In these tests we used dimensionless units.
\subsubsection{Gradient orientation test}\label{s:bborient}
Assuming $P_e= n_e T_e$, the Biermann battery term can be written as
\[
\frac{\partial\vec{B}}{\partial t} = \frac{c}{e}[\nabla T_e {\times}\nabla \ln(n_e)].
\]
Therefore, provided $\nabla T_e$ and $\nabla \ln(n_e)$ are constant,
one expects a steady and uniform production of magnetic field in
direction perpendicular to the plane defined by the above those two
gradients. Starting with the zero field initial conditions, no field
should occur when the two gradients are aligned and the field growth
should reach a maximum when the gradients are perpendicular.

\begin{figure}[htbp!]
  \begin{center}
$\!\!\!\!\!\!\!\!$\includegraphics[width=0.50\textwidth]{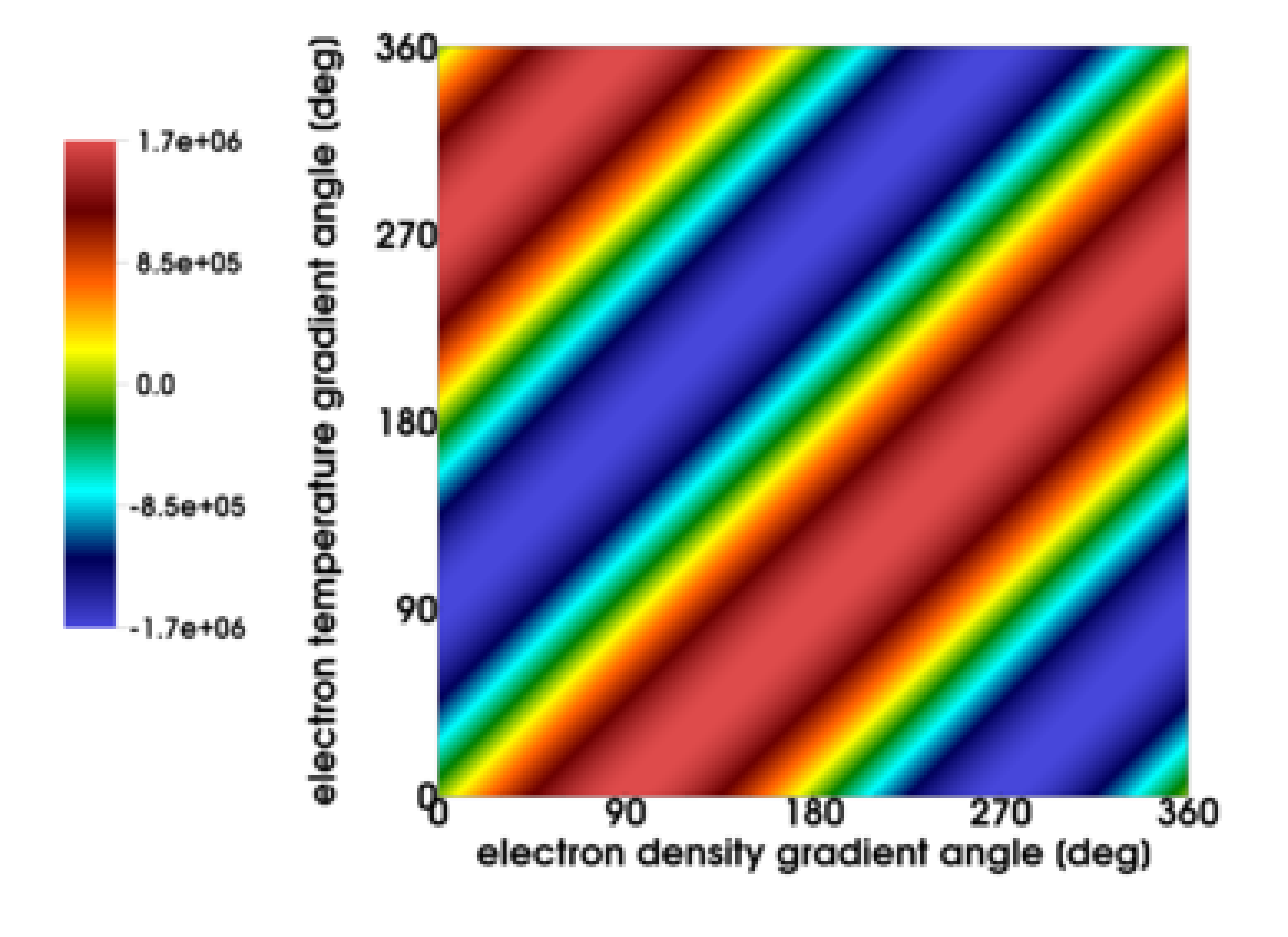}
    \caption{The value of magnetic field component, $B_z$, as a
      function of orientation of electron number density gradient and
      temperature gradient. The values of both gradients are
      constant. The generated magnetic field is strongest when the two
      gradients are orthogonal and no field is generated when they are
      parallel, as expected. See Sect.\ \ref{s:bborient} for details.}
    \label{f:BB360}
  \end{center}
\end{figure}

In Fig.~\ref{f:BB360}
we show the magnetic field component $B_z$, due to the Biermann
battery source term for various orientations of $\nabla \ln(n_e)$ and
$\nabla T_e$. The 2-D Cartesian domain covers a square region,
$(x,y)\in [0,360]\times[0,360]$. The orientation of $\nabla \ln(n_e)$
(not shown) varies from 0 to 360 degrees with the $x$ axis, while the
orientation of $\nabla T_e$ (not shown) changes from 0 to 360 degrees
along the $y$ axis, where 0 degrees angle corresponds to a gradient in
the direction of the $x$ axis. The magnitude of the temperature
gradient increases with $z$. The results shown in Fig.~\ref{f:BB360}
demonstrate that the generated magnetic field reaches a maximum when
the two gradients are orthogonal, and is zero when they are aligned,
as expected.
\subsubsection{Single-step Biermann battery test}\label{s:bbtoth}
One can compute the rate of magnetic field generation due to the
Biermann battery,
\[
\frac{\partial\vec{B}}{\partial
  t}=\nabla\times\frac{\nabla{P_e}}{n_e},
\]
by defining the initial distribution of the electron pressure and the
electron number density. If in addition, one assumes the medium is
initially static and there is no magnetic field, the problem
simplifies still further and one can verify the Biermann battery source term
implementation using essentially unmodified code.

\begin{figure*}[htbp!]
  \begin{center}
    \includegraphics[width=.32\textwidth]{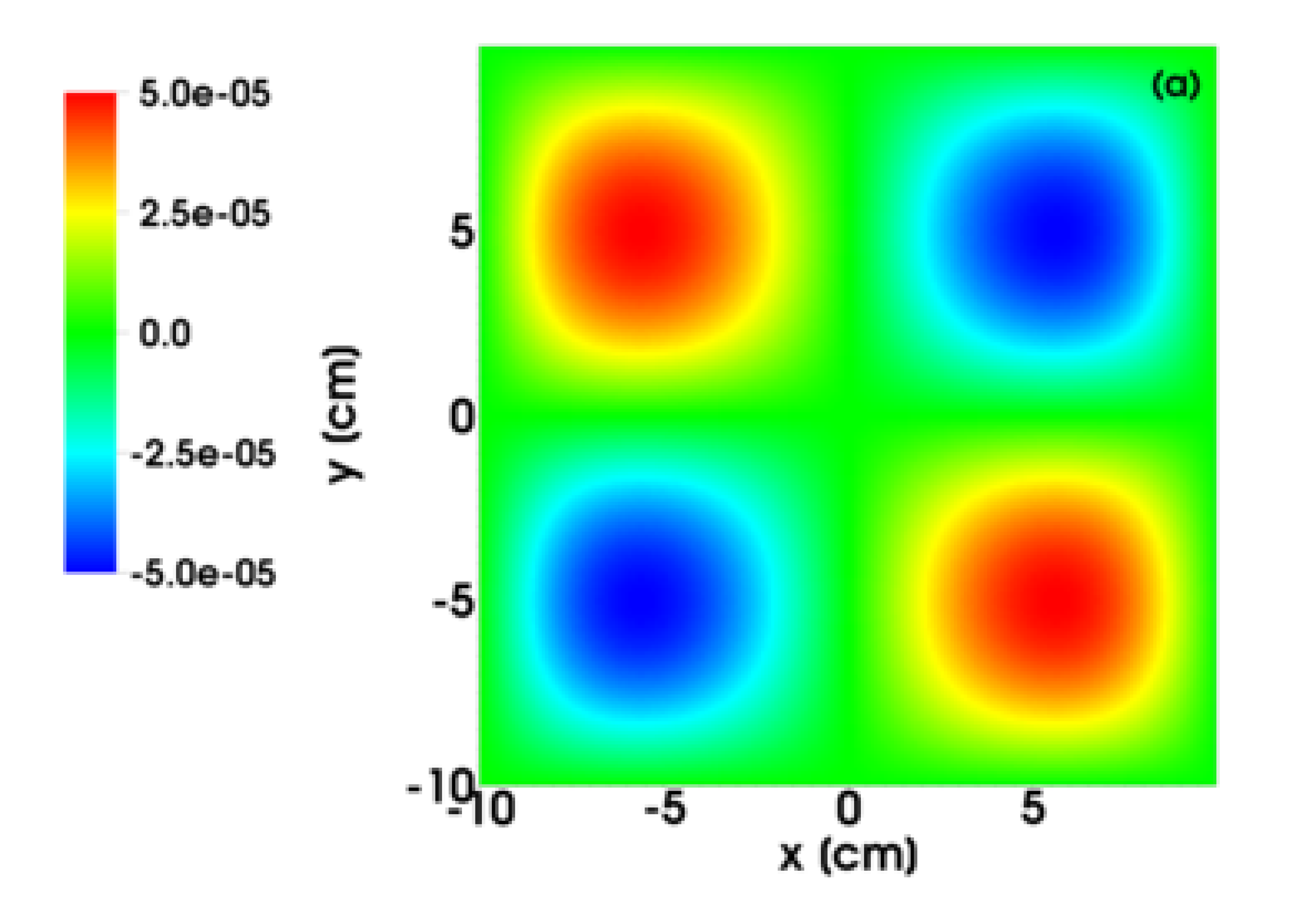}
    \includegraphics[width=.32\textwidth]{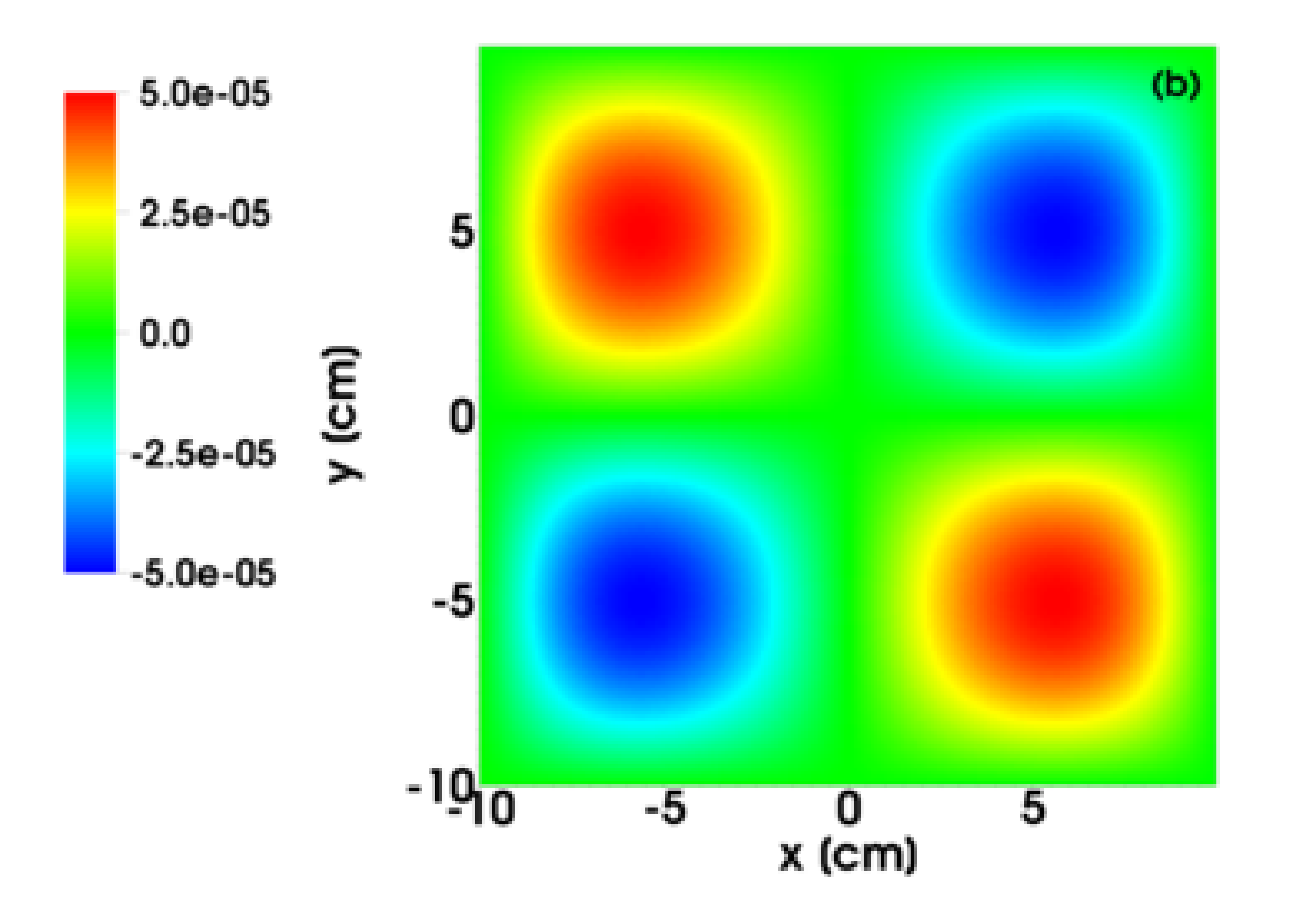}
    \includegraphics[width=.32\textwidth]{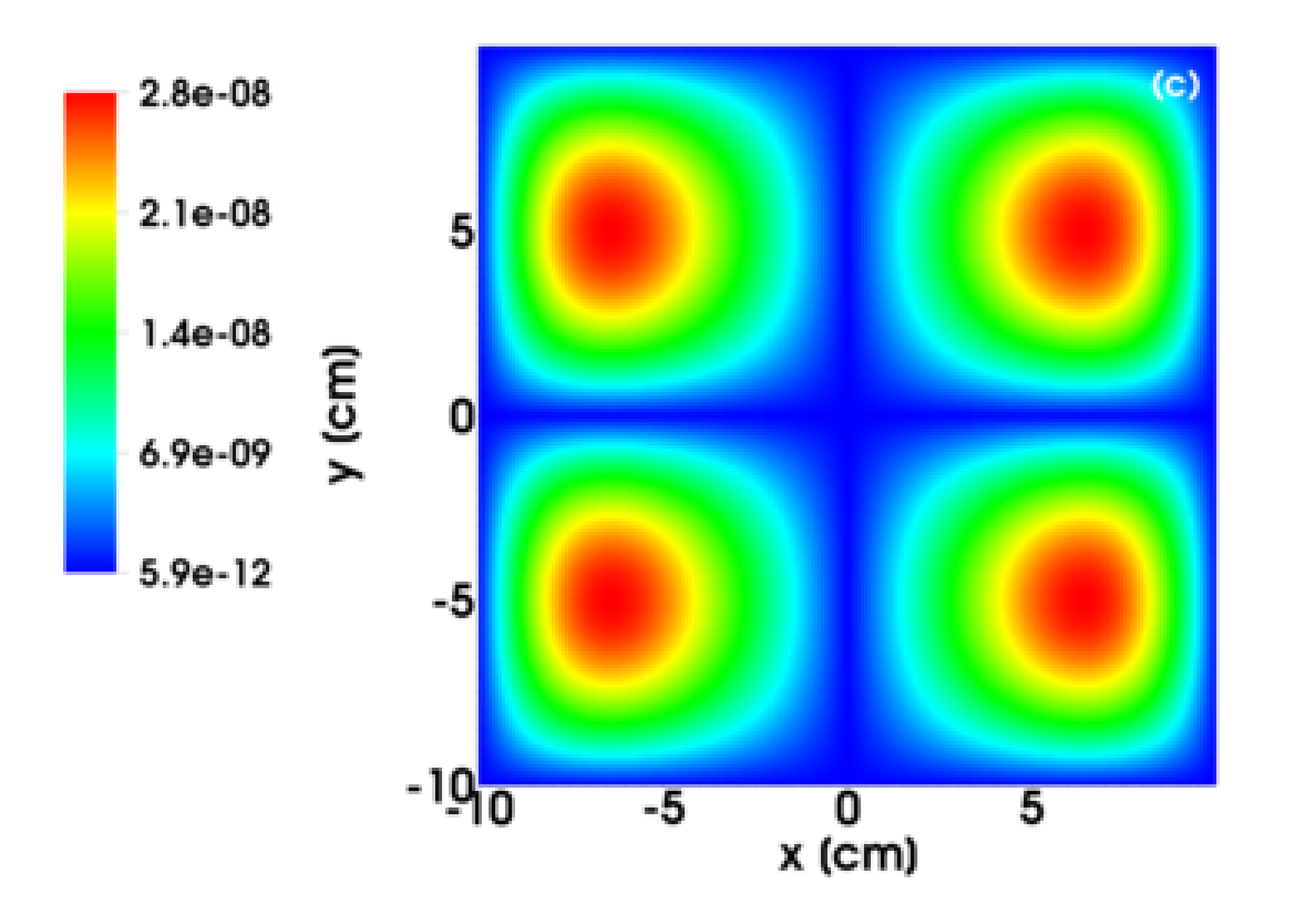}
    \caption{Solution verification of the \Proteus\ code Biermann
      battery magnetic field source term. The results are shown at $t
      = 0.05$ on the mesh $160{\times}160$ zones. (left panel)
      Computed solution; (center panel) exact solution; (right panel)
      absolute value of the difference between the exact and computed
      solution. Note the scale change between panels. See
      Sect.\ \ref{s:bbtoth} for details.}
    \label{f:BBTothAllFigs}
  \end{center}
\end{figure*}

We follow \citet{toth+12}, and assume the following initial conditions,
\begin{eqnarray*}
&& n_e=n_0+n_1\cos(k_xx),\\
&& P_e=P_0+P_1\cos(k_yy), \\
&& k_x=k_y=\frac{\pi}{10},
\end{eqnarray*}
with $n_0=p_0=1$ and $n_1=p_1=0.1$. For the above conditions, the
exact solution for the rate of magnetic field generation is,
\[
\frac{\partial {B_z}}{\partial t} = -\frac{k_xk_yn_1p_1\sin(k_xx)\sin(k_yy)}{[n_0+n_1\cos(k_xx)]^2}.
\]
Figure \ref{f:BBTothAllFigs}
shows the numerical and exact solutions, as well as the numerical
error. By performing this test using meshes with different resolutions,
we have confirmed that our implementation is second order accurate.
\subsubsection{Anisotropic conduction test}\label{s:condtest}
\begin{figure*}[htbp!]
  \begin{center}
    \includegraphics[height=0.35\textwidth,angle=0]{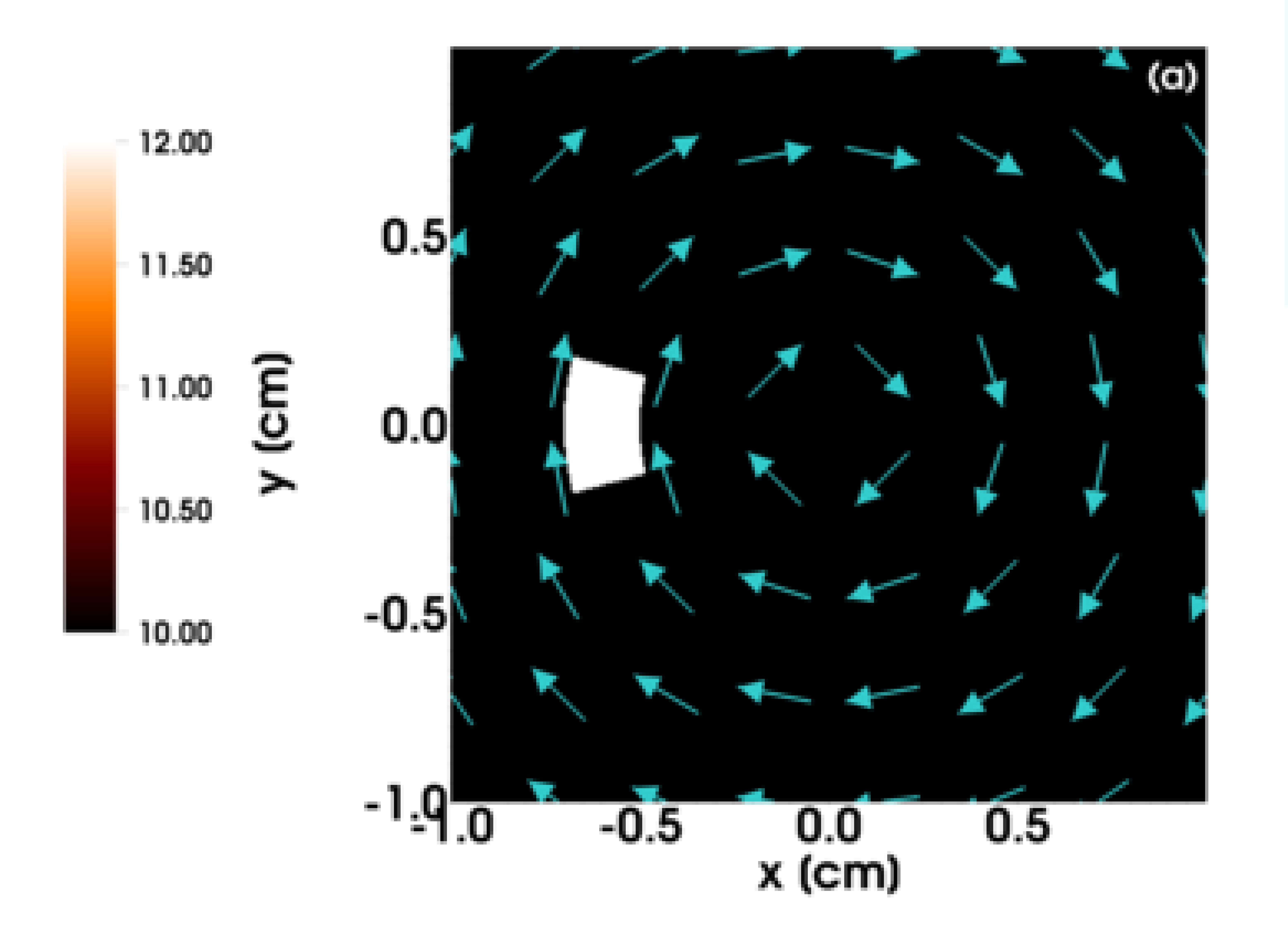}
    \includegraphics[height=0.35\textwidth,angle=0]{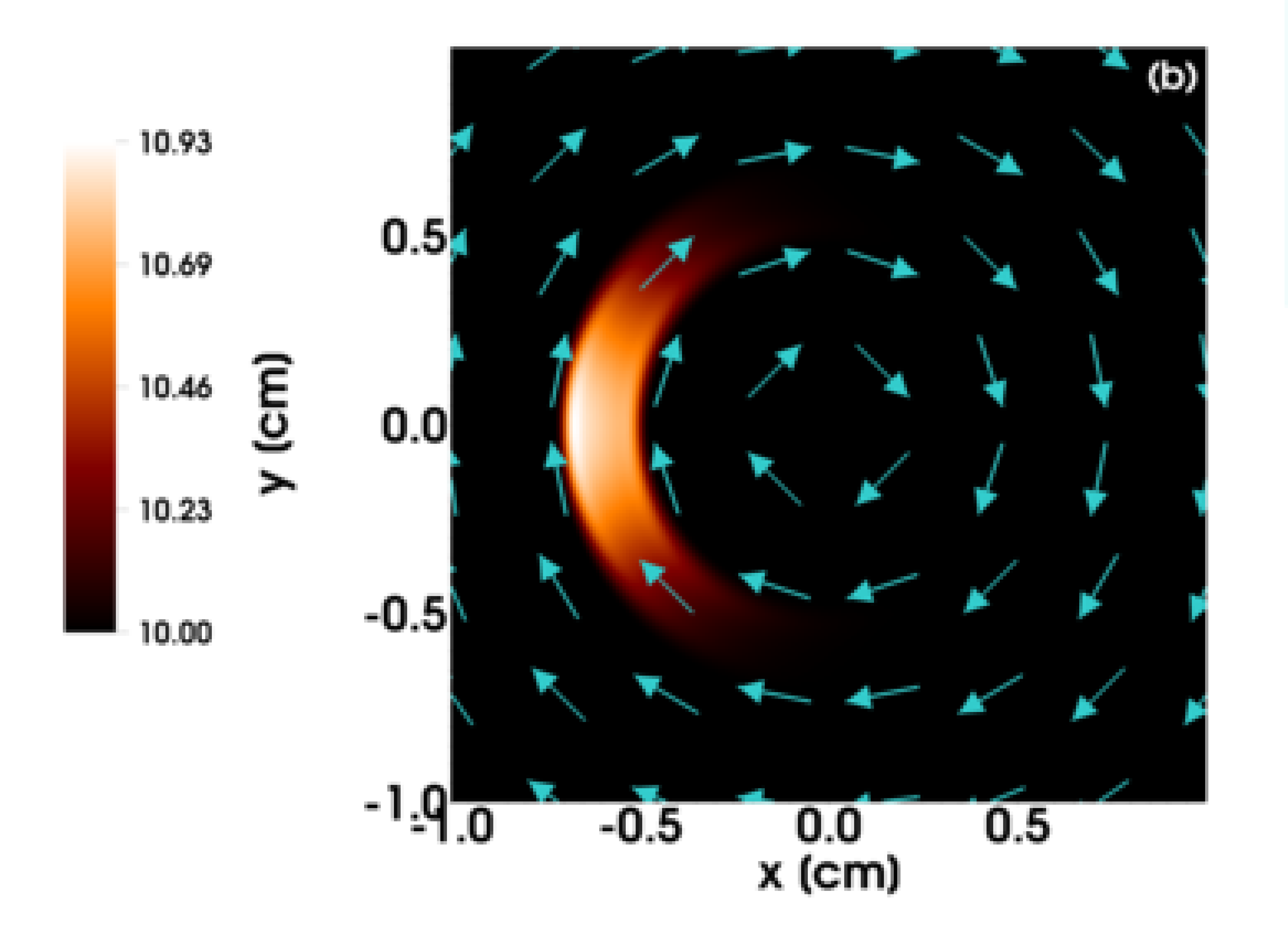}

    \includegraphics[height=0.35\textwidth,angle=0]{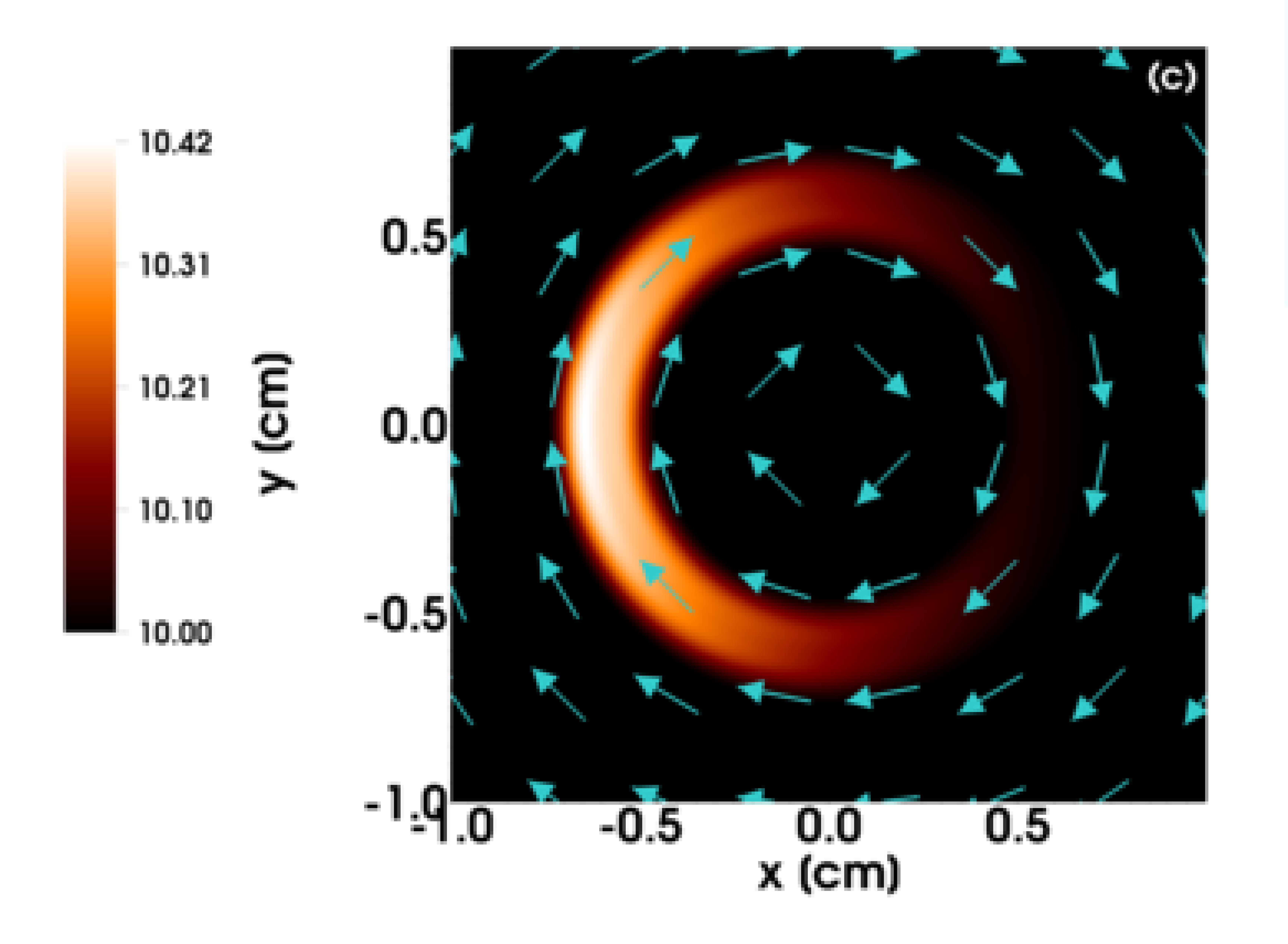}
    \includegraphics[height=0.35\textwidth,angle=0]{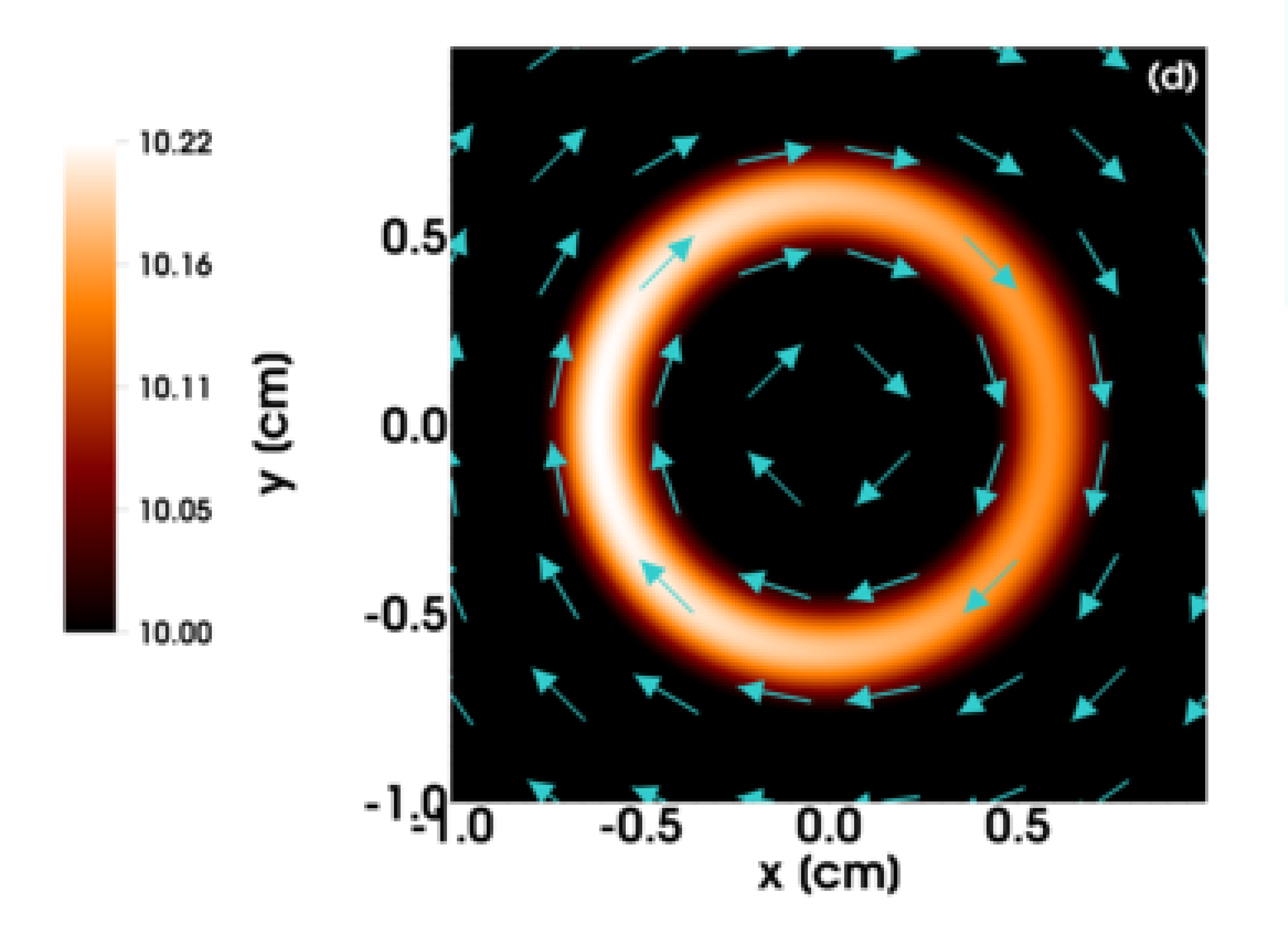}
    \caption{Solution verification of the \Proteus\ code anisotropic
      thermal conduction solver for the Parrish \& Stone
      \cite{parrish+05} test problem. The temperature
      distribution and magnetic field vectors are shown in the model
      with the mesh $400{\times}400$ zones. (a) t=0, (b) t=10, (c)
      t=50, (d) t=200. Heat transport is confined to a narrow ring
      bounded by magnetic field lines with only a small amount
      transported across the fields lines due to numerical diffusion,
      as expected. Note the scale changes between panels. See
      Sect.\ \ref{s:condtest} for details.}
    \label{f:ParrishStone}
  \end{center}
\end{figure*}

Figure \ref{f:ParrishStone}
%
%
shows the results of a test of anisotropic thermal conduction in a
circular magnetic field, as originally proposed by
\citet{parrish+05}. In this test, the computational domain is a
two-dimensional square region, $(x,y)\in [-1,1]\times[-1,1]$. The
initial conditions include a small region of higher temperature,
\[
T(r,\theta)=\left\{\begin{array}{ll}
        12 & \mbox{ if  $(0.5 \le r \le 0.7)$ and $(\frac{11}{12} \le \theta \le \frac{13}{12})$,} \\
        10 & \mbox{ otherwise.}
           \end{array}\right.
\]
In the ideal case and assuming heat cannot be transported across the
field lines, as the time proceeds, all heat should remain confined to
the annulus with the inner and outer radii of the small hot region. In
reality, however, one may expect that due to numerical errors some
heat will be transported across magnetic fields lines and diffuse
outside the ring.

In our test calculations, we set the parallel diffusion coefficient
$D_\|$ = 10 and $D_{\perp} = D_{\wedge} = 0$, where,
\begin{equation}
  \label{e:diffcoeff} 
D = \frac{\kappa}{\rho C_v},
\end{equation}
and $C_v$ is the specific heat at constant volume. For the above
choice of diffusion coefficients, and as we mentioned
earlier, the exact solution predicts that heat flows only within an
annulus defined by the magnetic field lines initially bounding the hot
region. Also, the temperature should be nearly evenly distributed
within the annulus by $t=200$.

\begin{figure}[htbp!]
  \begin{center}
    \includegraphics[width=0.50\textwidth,angle=0]{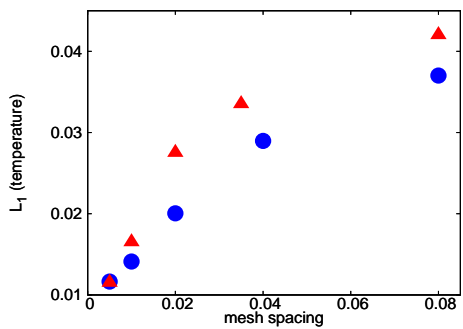}
    \caption{Convergence of the L1 error norm of temperature as a
      function of the mesh resolution for the Parish \& Stone
      \cite{parrish+05} test problem. Our results are shown with
      circles and those of Parish \& Stone with triangles.}
    \label{f:PSConvergence}
  \end{center}
\end{figure}

The residuals between simulation results and the exact solution for
this test problem are shown in Fig.~\ref{f:PSConvergence}
%
%
along with the results obtained by \citet{parrish+05} for
several different mesh resolutions. The temperature converges as the
mesh resolution increases, albeit slowly.

\citet{parrish+05} also studied a rate at which heat diffuses
across the magnetic field lines and estimated the corresponding
effective numerical perpendicular conduction coefficient,
$\kappa_\perp$. The smaller the value of this coefficient, the less
diffuse is the anisotropic thermal conduction solver.

\begin{figure}[htbp!]
  \begin{center}
    \includegraphics[width=0.50\textwidth,angle=0]{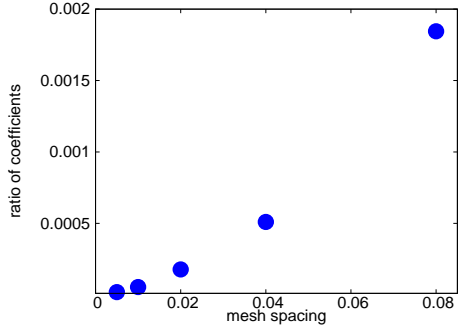}
    \caption{Dependence of the ratio of the effective numerical
      perpendicular conduction coefficient to parallel conduction
      coefficient, $\kappa_{\perp}$ / $\kappa_{\|}$ on the mesh
      resolution. See Sect.\ \ref{s:condtest} for discussion.}
    \label{f:PSPerp}
  \end{center}
\end{figure}

The value of the numerical conduction coefficient
characteristic of the \Proteus\ solver is shown in Fig.~\ref{f:PSPerp}
%
%
for different mesh resolutions. The results demonstrate the numerical
heat diffusion decreases with the resolution, as desired. We also note
that the performance of the \Proteus\ heat conduction module matches
that of the \citet{parrish+05} implementation.
\section{Application to Rayleigh-Taylor problem}\label{s:results}
In this section, we report the results of simulations of the
single-mode RTI in a basic plasma physics setting and study the
effects of thermal conduction and self-generated magnetic fields.
\subsection{Initial model}\label{s:ICs}
We considered a two-layer plasma system in two-dimensional Cartesian
geometry, $(x,y)$. In our initial model, the top plasma layer
consisted of high density carbon ($\rho=2$ g cm$^{-3}$) while the
bottom layer was made of a lower density carbon ($\rho=1$ g
cm$^{-3}$). The system was initially at rest and placed in a uniform
gravitational field, $\vec{g} = (0, -2{\times}10^{14})$ cm s$^{-2}$,
i.e.\ directed from the high density fluid to the low density fluid. This
configuration is RT-unstable.

We imposed a sinusoidal single-mode density perturbation at the
interface between the two layers, with peak-to-peak amplitude $A = 5$
$\mu$m and wavelength $\lambda = 71$ $\mu$m, as in
\citet{kuranz+10}. We defined the initial pressure profile by
integrating the equation of hydrostatic equilibrium with the minimum
ambient pressure of $5{\times}10^{11}$ dyn cm$^{-2}$. The resulting
temperature varied from about $3.8{\times}10^4$ K at the top of the
domain to approximately $1.3{\times}10^6$ K at the bottom.

In order to prevent the growth of spurious modes on small scales (due
to a finite mesh resolution), we used a diffuse interface with a
hyperbolic tangent density profile with 98\% of amplitude variation
over a distance of 5 $\mu$m. This procedure largely regularizes the
initial conditions and allows for more meaningful comparison of
simulation results obtained at different mesh resolutions. In the case
of diffuse interface, the RTI growth rate remains unchanged provided
the interface thickness is small compared to the perturbation
wavelength \cite{atzeni+04}. It is worth noting that in the HED
laser-driven RTI experiments some smearing of material interfaces is
expected to occur due to laser preheat \cite{kuranz+05}.

In application simulations, the mesh was uniformly resolved and
covered a rectangular region, $(x,y)\in [0,71]{\times}[0,284]$
$\mu$m. To assess convergence of the numerical solution, we varied the
number of mesh zones per the perturbation wavelength from 40 (L40
models) to 160 (L160 models). We used periodic boundary conditions in
the horizontal direction and enforced hydrostatic equilibrium at the
bottom and at the top of the computational domain.
\subsection{Simulation results}
\subsubsection{Pure hydrodynamics}\label{s:hyd}
\begin{figure}[htbp!]
  \begin{center}
    \includegraphics[height=0.405\textwidth,angle=0]{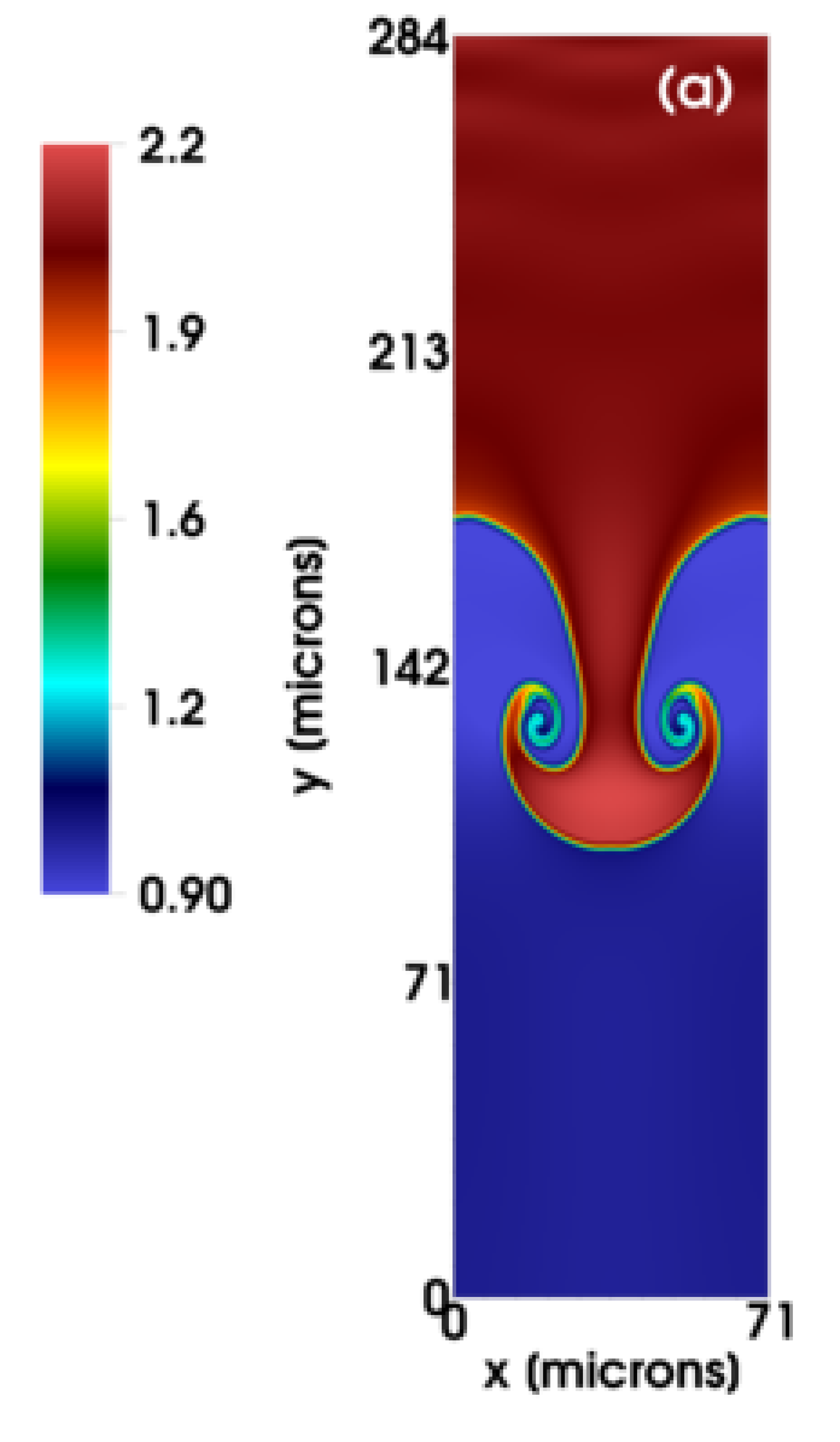}
    \includegraphics[height=0.405\textwidth,angle=0]{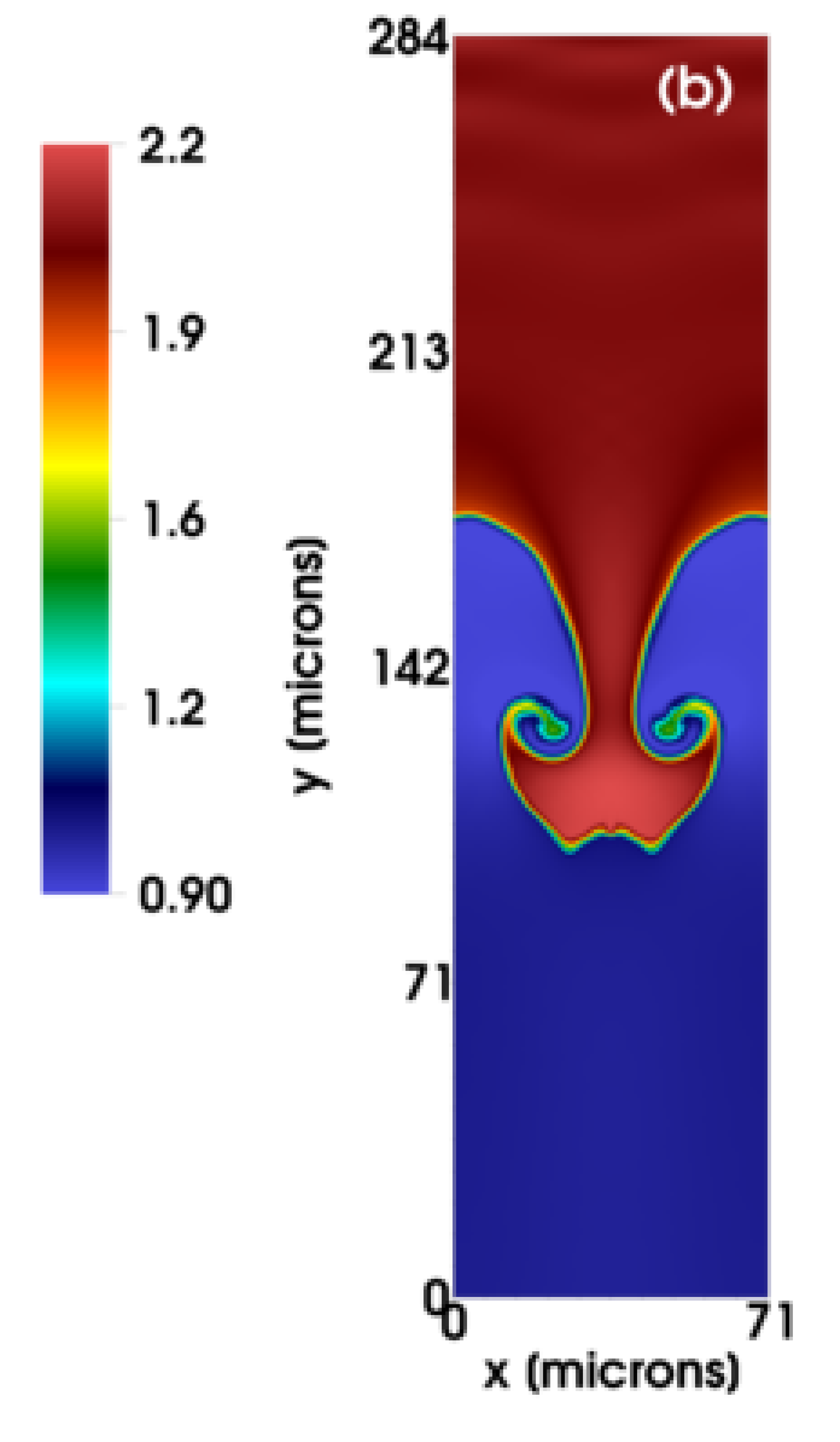}

    \includegraphics[height=0.405\textwidth,angle=0]{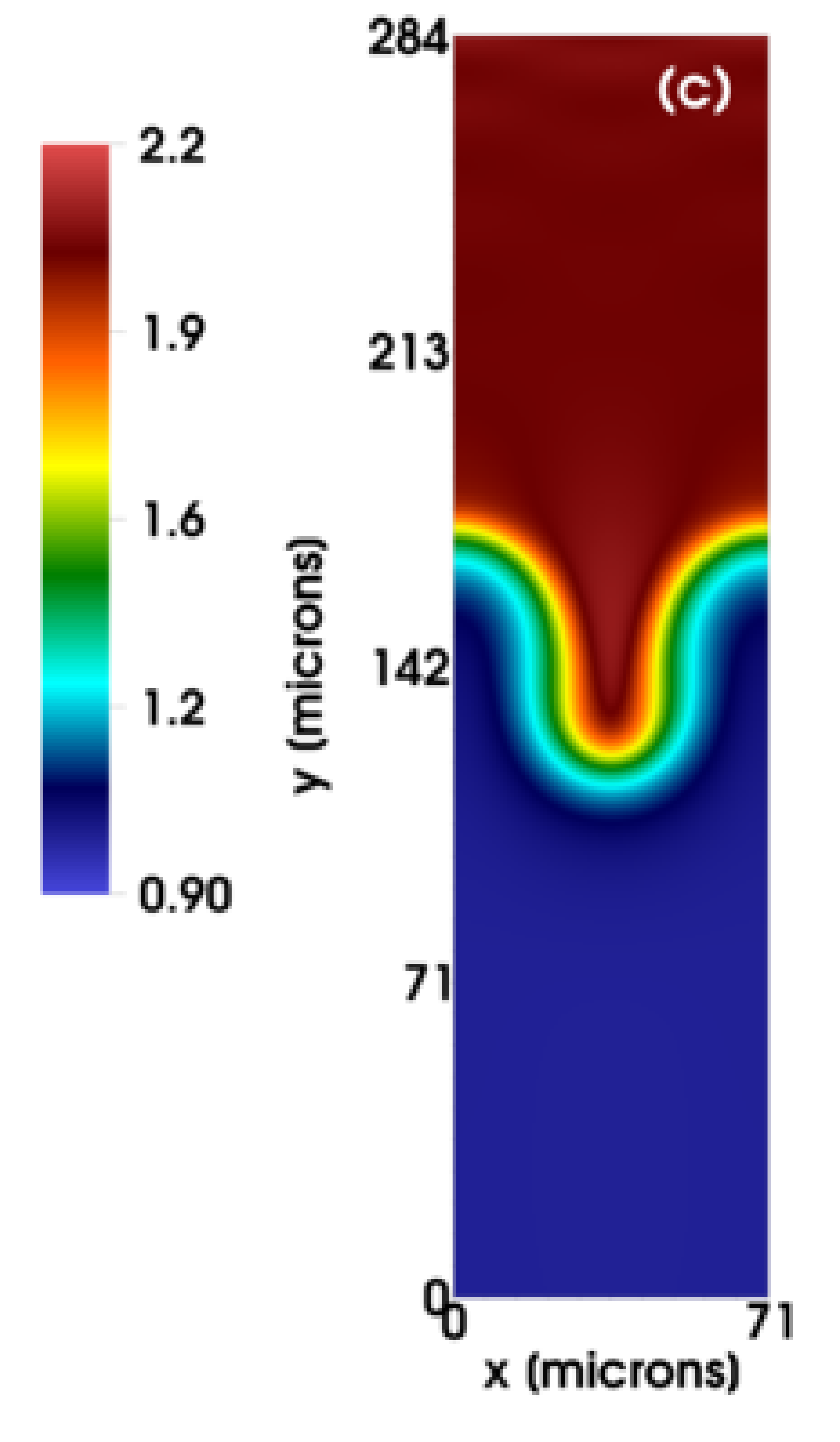}
    \includegraphics[height=0.405\textwidth,angle=0]{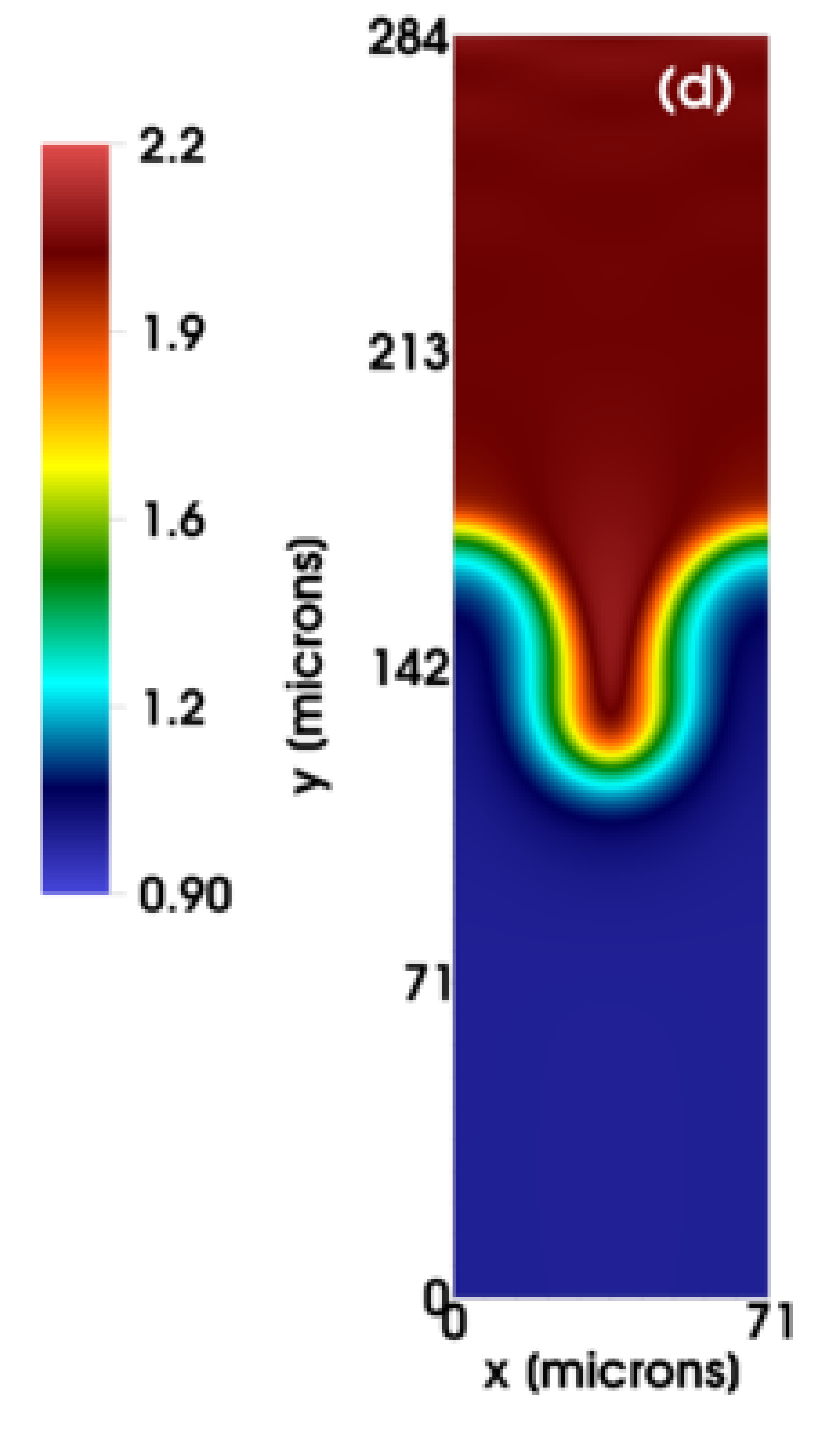}
    \caption{Rayleigh-Taylor instability: L160 model with various physics
      effects. Shown is the density distribution at $t=20$ ns. (a)
      Hydrodynamics only. (b) Hydrodynamics with self-generation of
      magnetic fields. (c) Hydrodynamics with thermal conduction. (d)
      Hydrodynamics with self-generation of magnetic fields and thermal
      conduction.}
    \label{f:density20ns}
  \end{center}
\end{figure}

\begin{figure}[htbp!]
  \begin{center}
    \includegraphics[height=0.405\textwidth,angle=0]{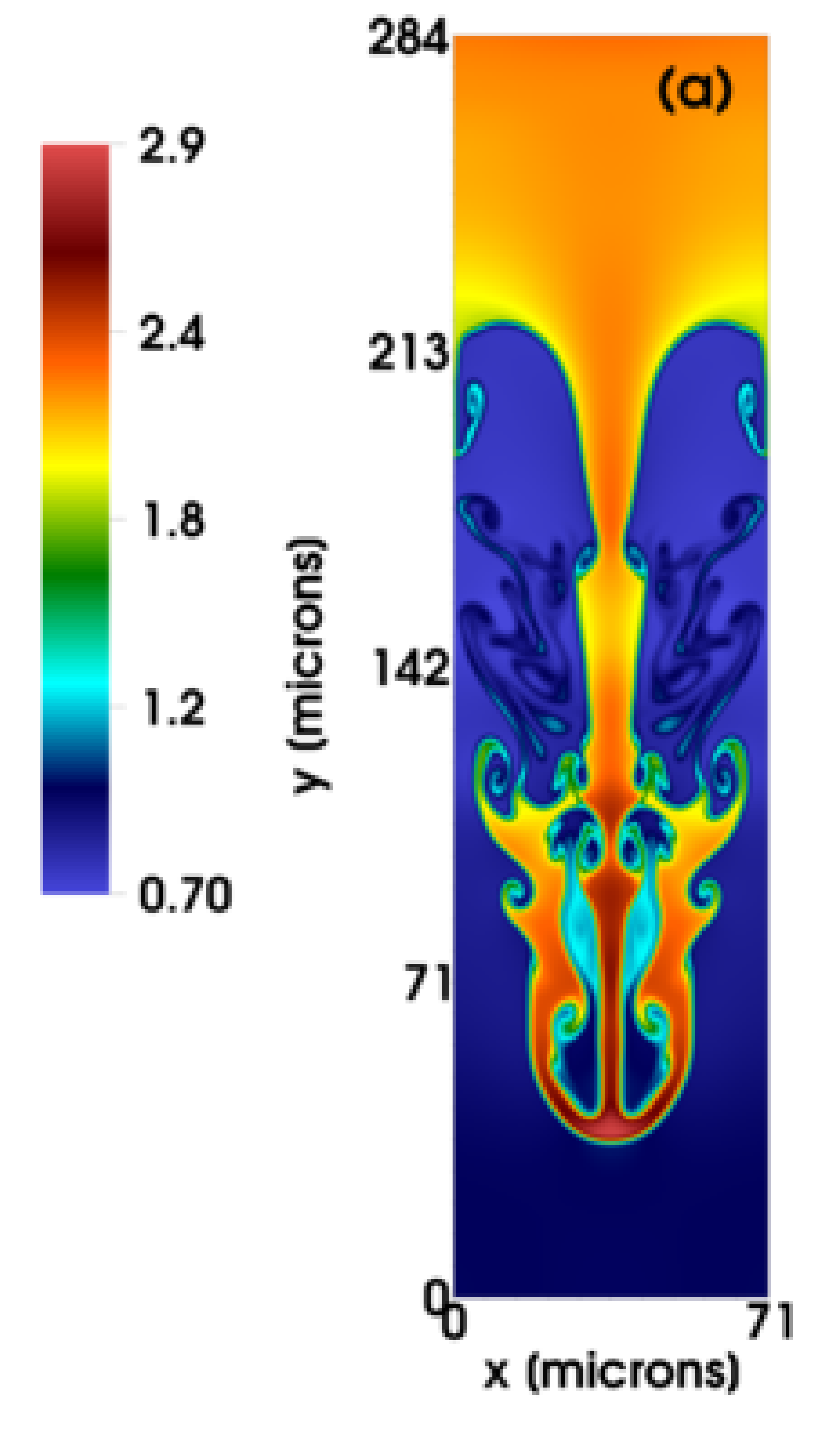}
    \includegraphics[height=0.405\textwidth,angle=0]{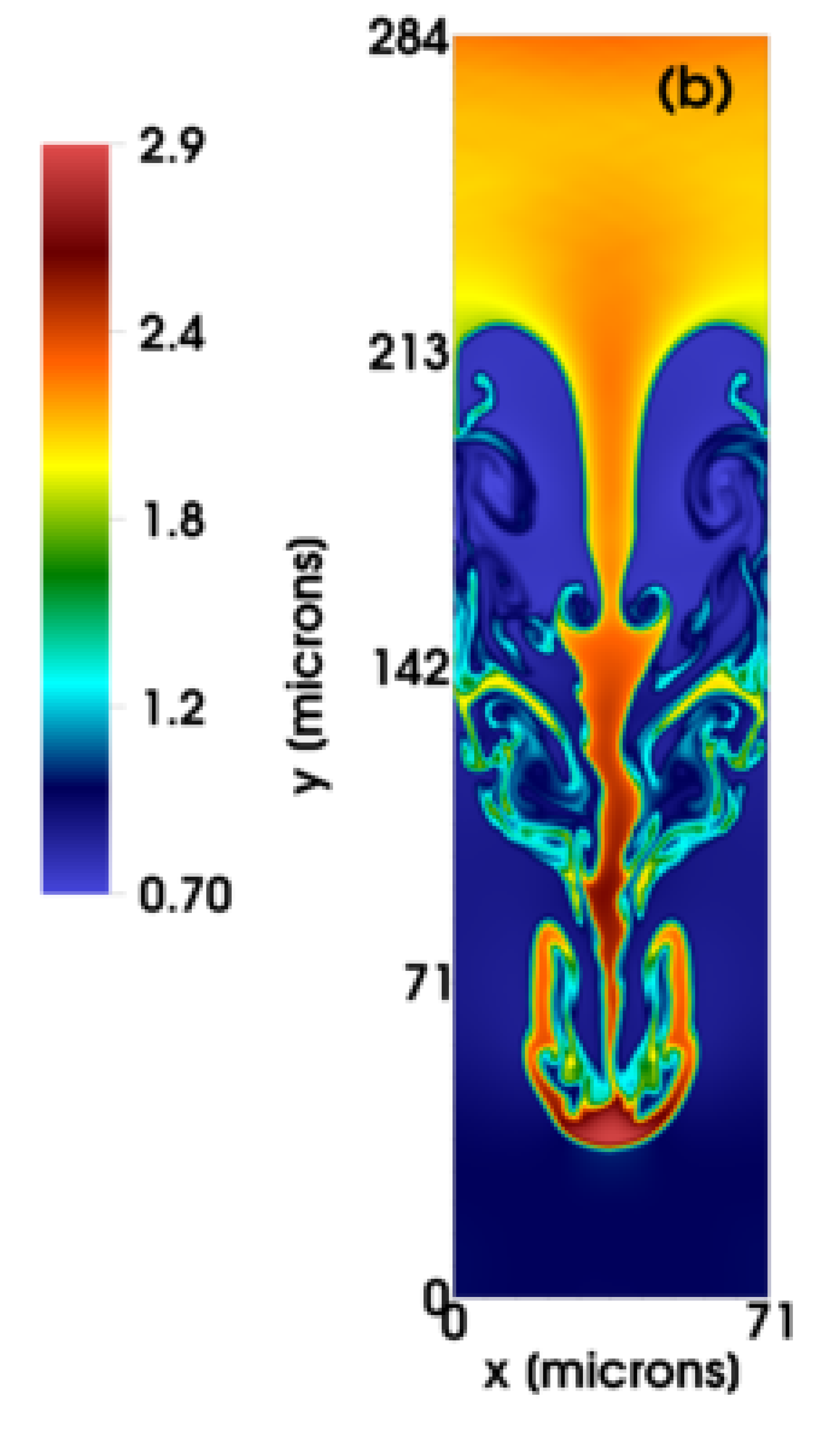}

    \includegraphics[height=0.405\textwidth,angle=0]{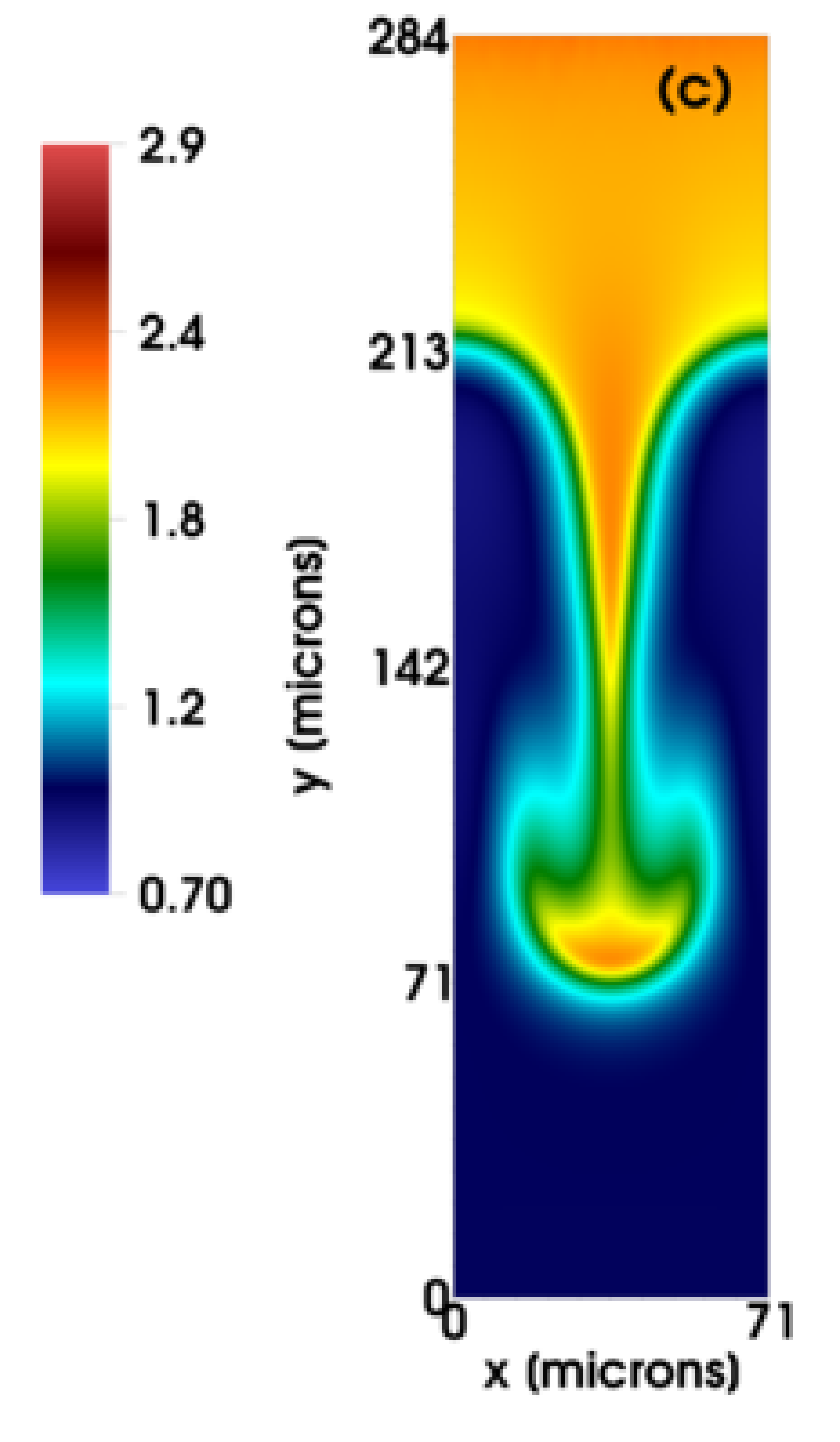}
    \includegraphics[height=0.405\textwidth,angle=0]{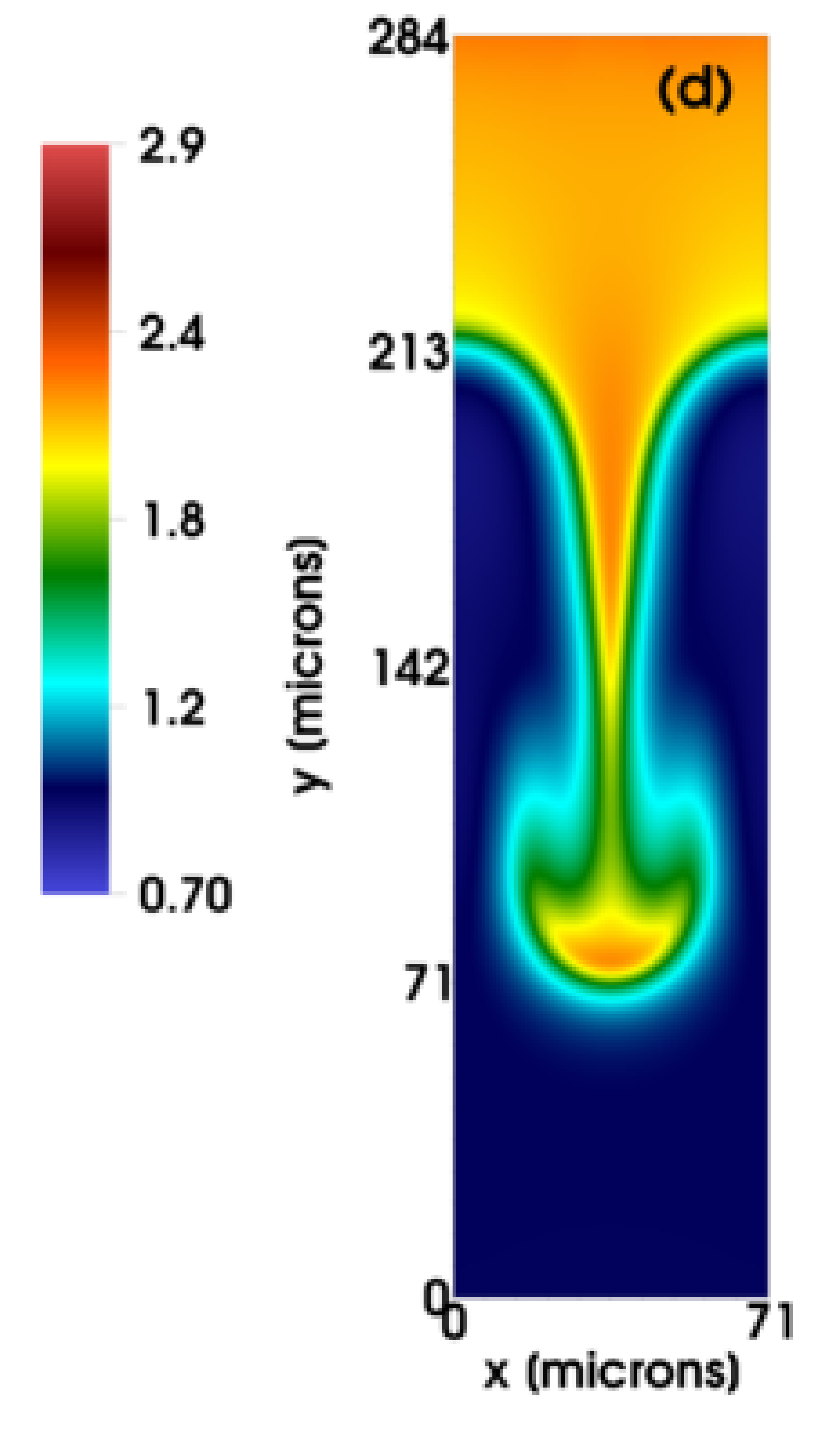}
    \caption{Rayleigh-Taylor instability: L160 model with various physics
      effects. Shown is the density distribution at $t=35$ ns. (a)
      Hydrodynamics only. (b) Hydrodynamics with self-generation of
      magnetic fields. (c) Hydrodynamics with thermal conduction. (d)
      Hydrodynamics with self-generation of magnetic fields and thermal
      conduction.}
    \label{f:density35ns}
  \end{center}
\end{figure}

In this model, the initial perturbation grows and retains its general
sinusoidal profile on large scale until $t\approx 15$ ns. Soon after
that time, the fluid on the spike shoulders begins to roll up
resulting in the familiar mushroom-like RTI morphology
(Fig.~\ref{f:density20ns}a).
%
%
At $t \approx 25$ ns, the initial vortex is shed and a new one is
formed at the leading section of the spike. The vorticity associated
with those large scale flow features eventually induces a complex
mixed flow structure on small scales dominating the mid-section of the
spike at the final time (see Fig.~\ref{f:density35ns}a).

\begin{figure*}[htbp!]
  \begin{center}
    \includegraphics[height=0.35\textwidth,angle=0]{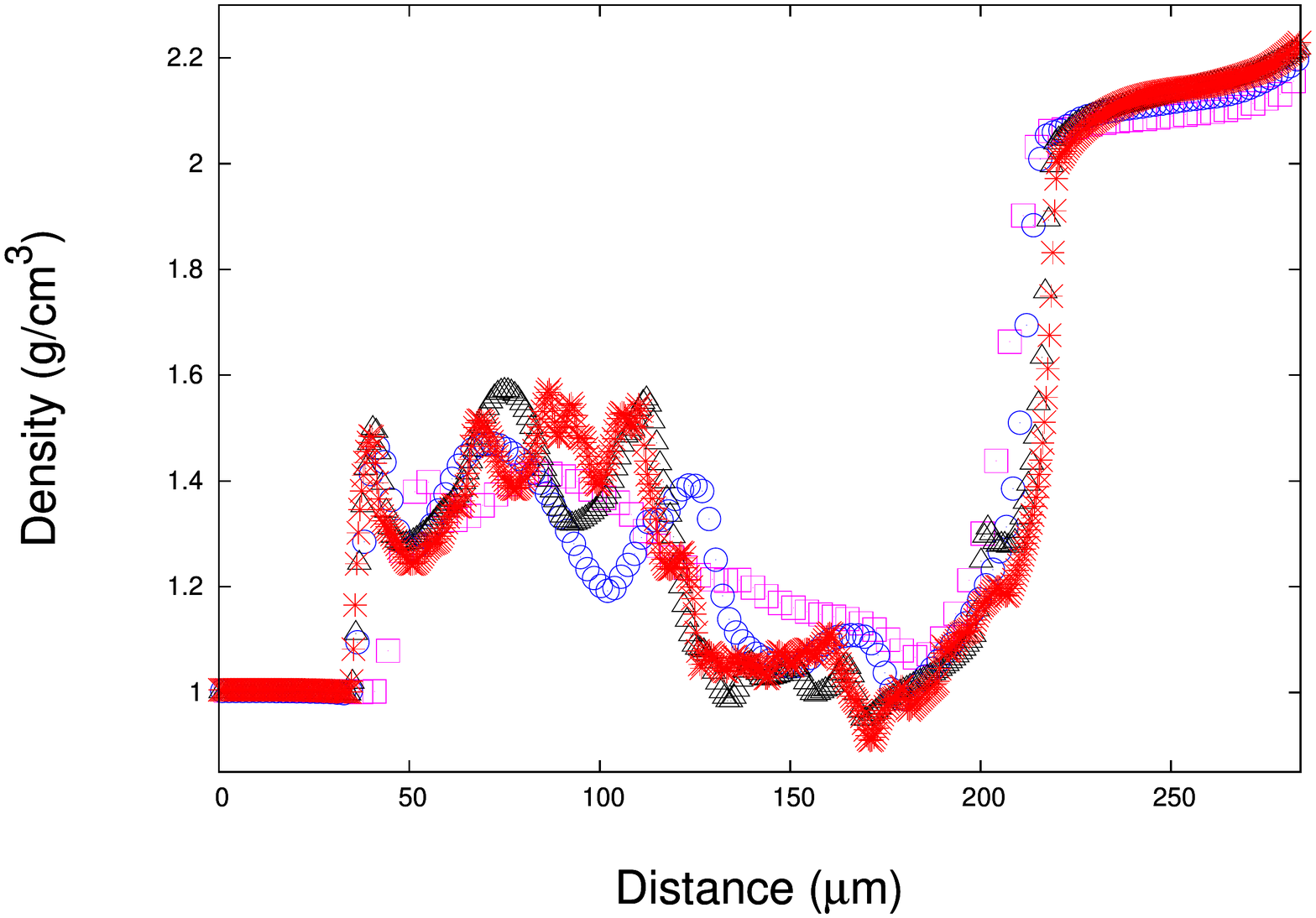}
    \includegraphics[height=0.35\textwidth,angle=0]{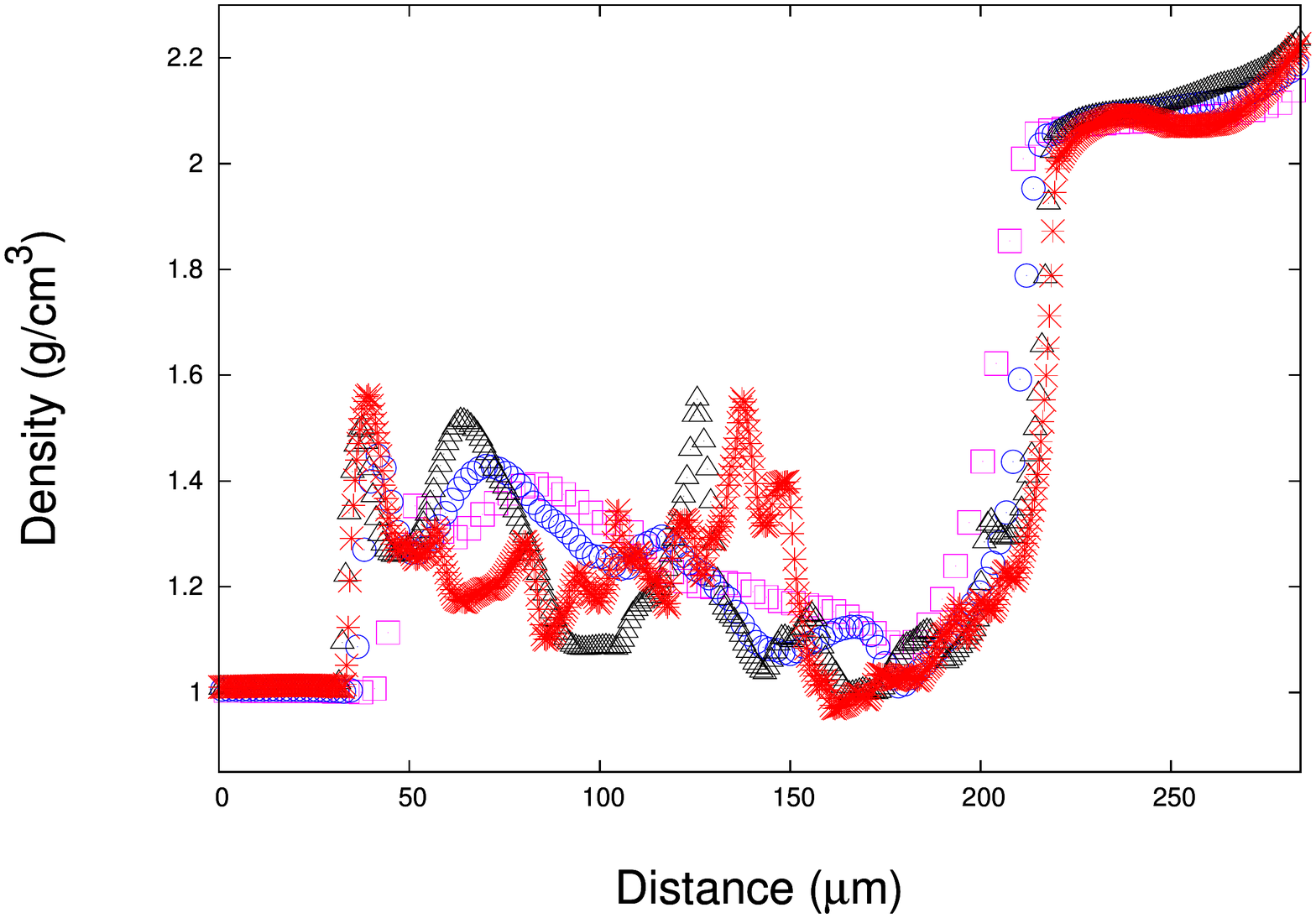}

    \includegraphics[height=0.35\textwidth,angle=0]{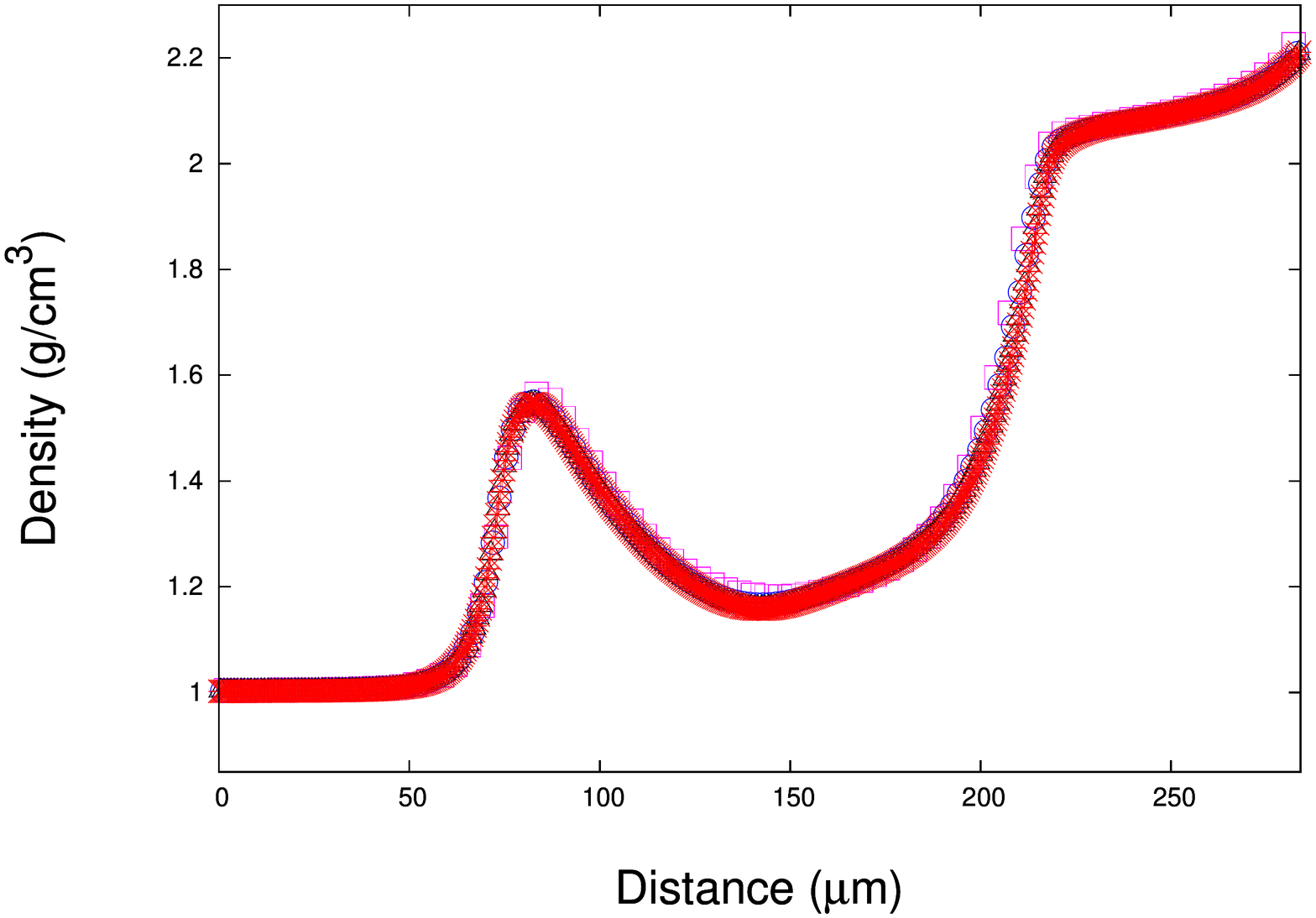}
    \includegraphics[height=0.35\textwidth,angle=0]{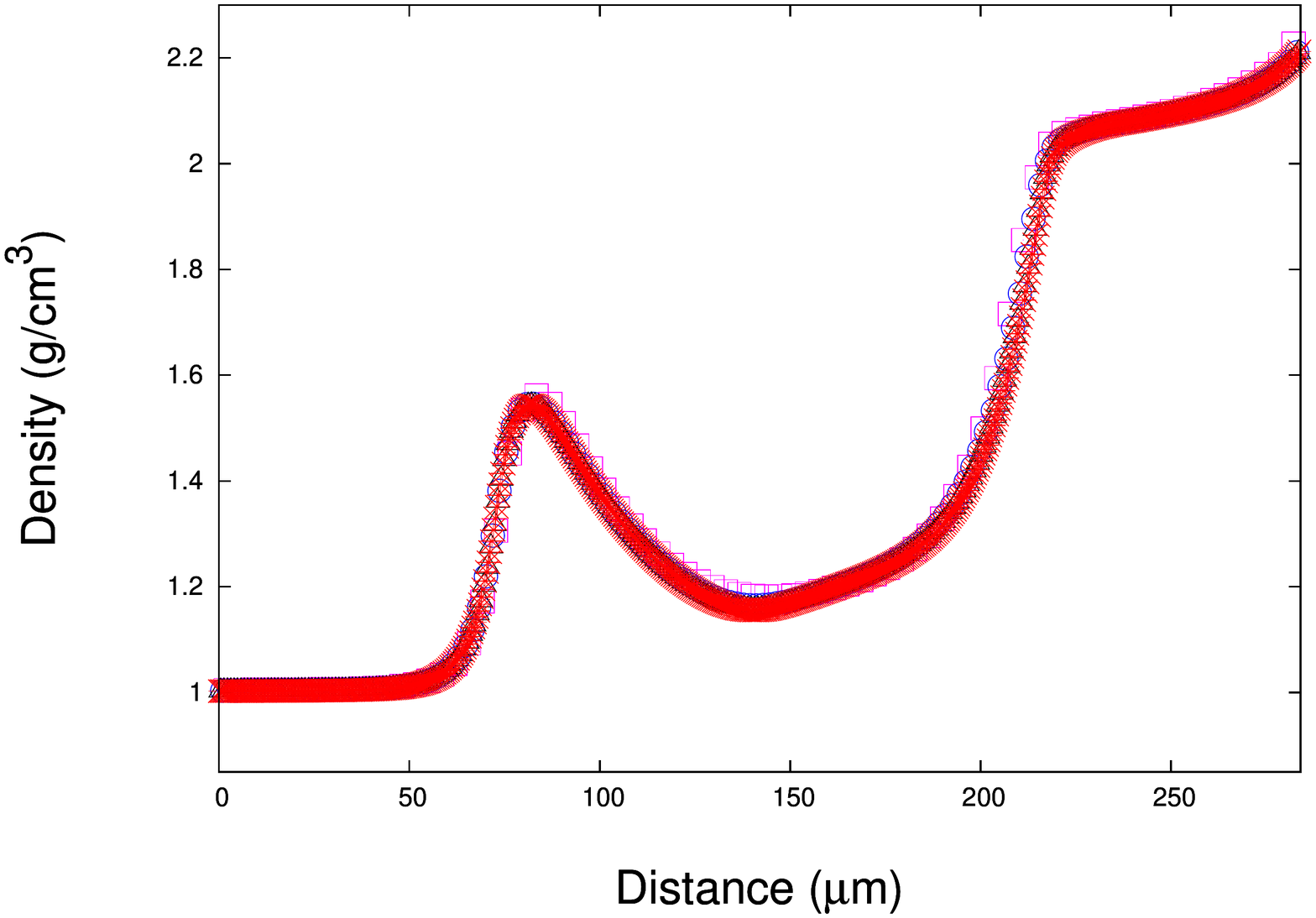}
    \caption{Lateral averages of density as a function of vertical
      position for L20 (squares), L40 (circles), L80 (triangles), and
      L160 (asterisks) models at the final time. (a) Hydrodynamics
      only. (b) Hydrodynamics with self-generation of magnetic
      fields. (c) Hydrodynamics with thermal conduction. (d)
      Hydrodynamics with self-generation of magnetic fields and
      thermal conduction.}
    \label{f:avps}
  \end{center}
\end{figure*}

Figure \ref{f:avps}a shows lateral (in the direction perpendicular to
gravity) averages of density in a model with hydrodynamics only at the
final time for various mesh resolutions. In this case, positions of
spike and bubble tips converge already at L40. The lateral density
averages also appear to converge to a single distribution inside the
mixed region (e.g.\ low density region around 55 $\mu$m and a broad
low density area around 150 $\mu$m). These results are qualitatively
consistent with the trends reported by \citet{budde+10}.
%
%
\subsubsection{Hydrodynamics with self-generated magnetic field}\label{s:hydmag}
\begin{figure*}[htbp!]
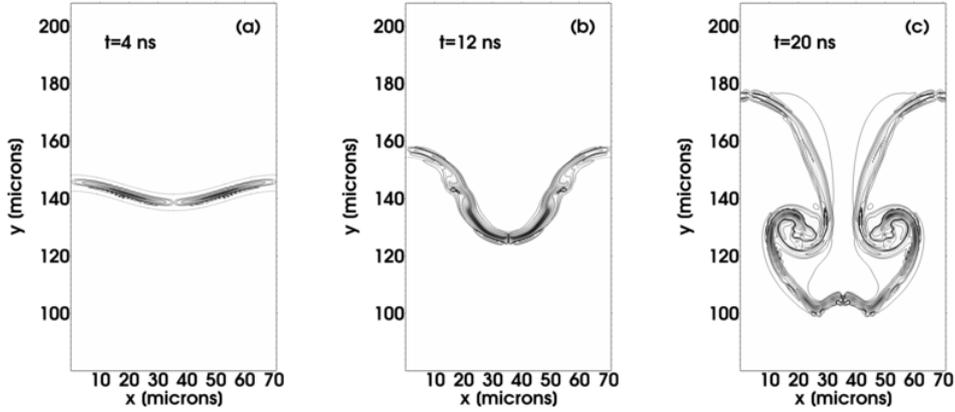

  \begin{center}
    \includegraphics[width=.32\textwidth,trim=23cm 10cm 23cm 5cm,clip=true]{f9_left.pdf}
    \includegraphics[width=.32\textwidth,trim=23cm 10cm 23cm 5cm,clip=true]{f9_center.pdf}
    \includegraphics[width=.32\textwidth,trim=23cm 10cm 23cm 5cm,clip=true]{f9_right.pdf}
    \caption{Development of the Rayleigh-Taylor instability in the
      non-conducting model with self-generation of magnetic
      fields. (a) $t=4$ ns; (b) $t=12$ ns; (c) $t=20$ ns.  The model
      resolution is L160. Boundaries of the interface region between
      light and heavy fluid are marked with contours of mass fraction
      of the heavy fluid of 0.01 and 0.99 (dashed lines). Contours of
      the absolute value of magnetic field component, $|B_z|$, are
      shown with solid lines. The magnetic field contours start at
      $\log|B_z| = 4.5$ and are spaced by 0.5 dex. The magnetic
      field strength is additionally color-coded in log-scale; the
      darker the color the stronger the field. See
      Sect.\ \ref{s:hydmag} for discussion.}
    \label{f:HydroMag_L160_field_mix}
  \end{center}
\end{figure*}

Compared to the pure hydro model, the influence of self-generated
magnetic fields on RTI appears negligible during the linear growth
phase (Fig.\ \ref{f:HydroMag_L160_field_mix}a).
%
%
However, around the time when the width of mixed layer reaches about
50\% of the perturbation wavelength, two small bulging structures
become clearly visible at the interface located roughly half way
between the spike tip and the bubble tip
(Fig.\ \ref{f:HydroMag_L160_field_mix}b). We identify this new
flow feature as an additional RTI mode. This mode is observed only in
models with self-generated magnetic fields and without conduction (see
Sect.\ \ref{s:hydcondmag} below for a description of the corresponding
model with conduction).

\begin{figure}[htbp!]
  \begin{center}
    \includegraphics[width=0.55\textwidth,trim=0.5cm 7cm 0.5cm 5cm,clip=true]{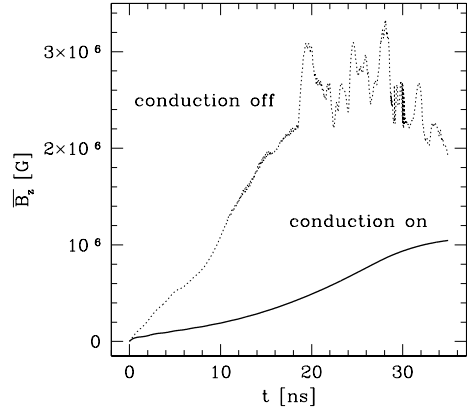}
    \caption{Time evolution of the maximum horizontally averaged
      magnetic field strength in L160 RTI models with
      self-magnetization. (dotted) without conduction; (solid) with
      conduction.}
    \label{f:maxaveB}
  \end{center}
\end{figure}

\begin{figure*}[htbp!]
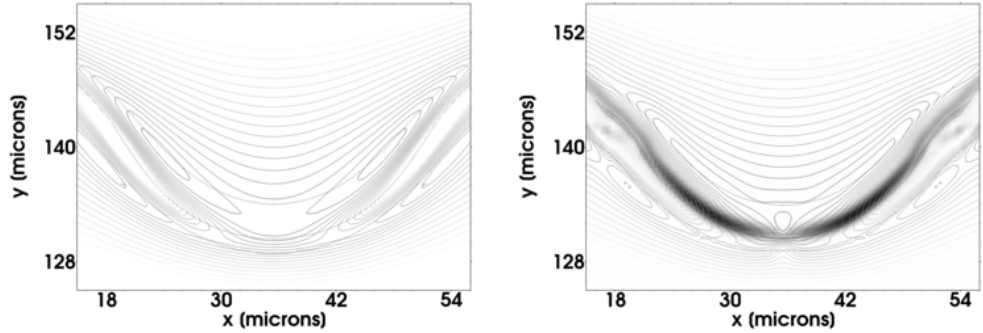

  \begin{center}
    \includegraphics[width=.49\textwidth,trim=3cm 15cm 3cm 5cm,clip=true]{f11_left.pdf}
    \includegraphics[width=.49\textwidth,trim=3cm 15cm 3cm 5cm,clip=true]{f11_right.pdf}
    \caption{Structure of the flow near the tip of RT spike at $t=10$ in
      non-conducting models (a) without and (b) with self-generation
      of magnetic field.  The resolution of both models is
      L160. Boundaries of the interface region between light and heavy
      fluid are marked with contours of mass fraction of the heavy
      fluid of 0.01, 0.5, and 0.99 (dashed lines). The magnitude of
      fluid velocity is shown with 20 equispaced contour lines between
      $1{\times}10^5$ cm/s and $2.5{\times}10^5$ cm/s. The darker the
      contour line the higher the velocity. See Sect.\ \ref{s:hydmag} for
      discussion.}
    \label{f:HydroVsHydroMag}
  \end{center}
\end{figure*}

During that time, the magnetic field rapidly builds up inside the
interface with the average field reaching $1$ MG around $t=10$ ns (dotted
line in Fig.\ \ref{f:maxaveB}).
At that time, the maximum field strength is $B_z\approx{3.7}$ MG
(plasma $\beta{\approx}11$) along leading sides of the sinking
spike. This additional pressure modifies the flow dynamics in the
interface region with the most profound effect occurring near the
spike tip. Figure \ref{f:HydroVsHydroMag}
allows for comparison of the flow structure in the current model
(right panel) to that in the pure hydro model (left panel). The
important difference between the two solutions is a contribution of
the self-generated magnetic field to the plasma pressure. The ratio
between magnetic pressure and the gas pressure (inverse of plasma
$\beta$) is shown as a gray contour map. The maximum pressure ratio is
reached very near the middle of the interface (second from the top
dashed contour line corresponds to mass fraction of the heavy fluid of
50\%) and closer to the spike tip. The contribution due to magnetic
pressure then gradually decreases away from that point and reaches
zero at the very spike tip (no field is generated in that area). The
fluid elements close to the spike center line are no longer smoothly
accelerated (spacing between the velocity contours gradually increases
until the mid-center of the interface profile in the left panel in
Fig.\ \ref{f:HydroVsHydroMag}), but are deflected toward the center
line by pressure gradients due to generated magnetic field. In the
process, the flow ``chokes'' with fluid elements located closest to
the spike's center line being decelerated most. This process of
slowing down the flow near the spike's center becomes increasingly
more efficient as the magnetic field grows. Figure
\ref{f:HydroMag_L160_field_mix}c (and also Fig.\ \ref{f:density20ns}b)
shows the central section of the spike surface bent inward by about
4 $\mu$m ($\approx 5$\% of the mixing length width) over the course
of $10$ ns.

The rapid initial growth of the field saturates around $t=20$ ns with
the average maximum field strength reaching approximately $2.5$ MG
(dotted line in Fig.\ \ref{f:maxaveB}). At those intermediate times,
$\beta$ is typically $< 1$ inside the interface and the strongest
fields are observed at the outer edges of the RT mushroom cap. The
rapid increase of the maximum field strength that starts around
$t\approx 18$ ns also occurs in those regions and is due to fast
bending of the magnetic field lines when the bottom sections of the
cap roll inward.  Strong fluctuations of the magnetic field observed
at later times occur on small scales (see below) and their amplitude
will likely be smaller in 3-D models and in realistic situations when the
symmetry of the flow is broken by perturbations.
\subsubsection{Hydrodynamics with thermal conductivity}\label{s:hydcond}
The large temperature gradients present in the initial conditions make
thermal conductivity a potentially important physical effect in our RTI
system. Indeed, the most striking and common feature of all
thermally conducting models considered here is strong mass diffusion
across the interface, the resulting dramatically reduced amount of
density structure, and much smoother morphology of the flow. In
particular, the RT mushroom cap seen in the pure hydro model is much
smoother when conduction is accounted for, and we do not observe any
significant density roll-up (see Fig.\ \ref{f:density20ns}c).
\subsubsection{Hydrodynamics with thermal conductivity and self-generated magnetic field}\label{s:hydcondmag}
The addition of self-generation of magnetic fields does not
qualitatively change the evolution of thermally conducting model
discussed in the previous section. The interface suffers strong
diffusion at early times (Fig.~\ref{f:density20ns}d) and the RT
mushroom cap remains underdeveloped (Fig.~\ref{f:density35ns}d). The
initial magnetic field growth rate is by a factor of about 5 smaller
than in the non-conducting case (solid line in Fig.\ \ref{f:maxaveB}),
and the field strength reaches approximately $190$ kG at $t=10$
ns. Moreover, we do not see enhancement of small scale structure
found in the non-conducting self-magnetized model
(Fig.~\ref{f:density35ns}b). This is expected given the strong
diffusive effects due to thermal conductivity. In consequence, rapid
fluctuations of the magnetic field are absent in this case.
\section{Discussion}
\subsection{Comparison of physics effects on RTI mixing}
\subsubsection{Mixing on large scales}\label{s:lsmixing}
To quantify the effects of mixing on large scales, we introduced a
scalar passively advected with the plasma in the simulations. At the
initial time, the value of the passive scalar was set to one inside
the dense layer and to zero inside the light material.  As the
simulation progressed, the tracer was advected and tracked
distribution of the dense carbon.

\begin{figure*}[htbp!]
  \begin{center}
    \includegraphics[width=0.35\textwidth,angle=0,clip=true]{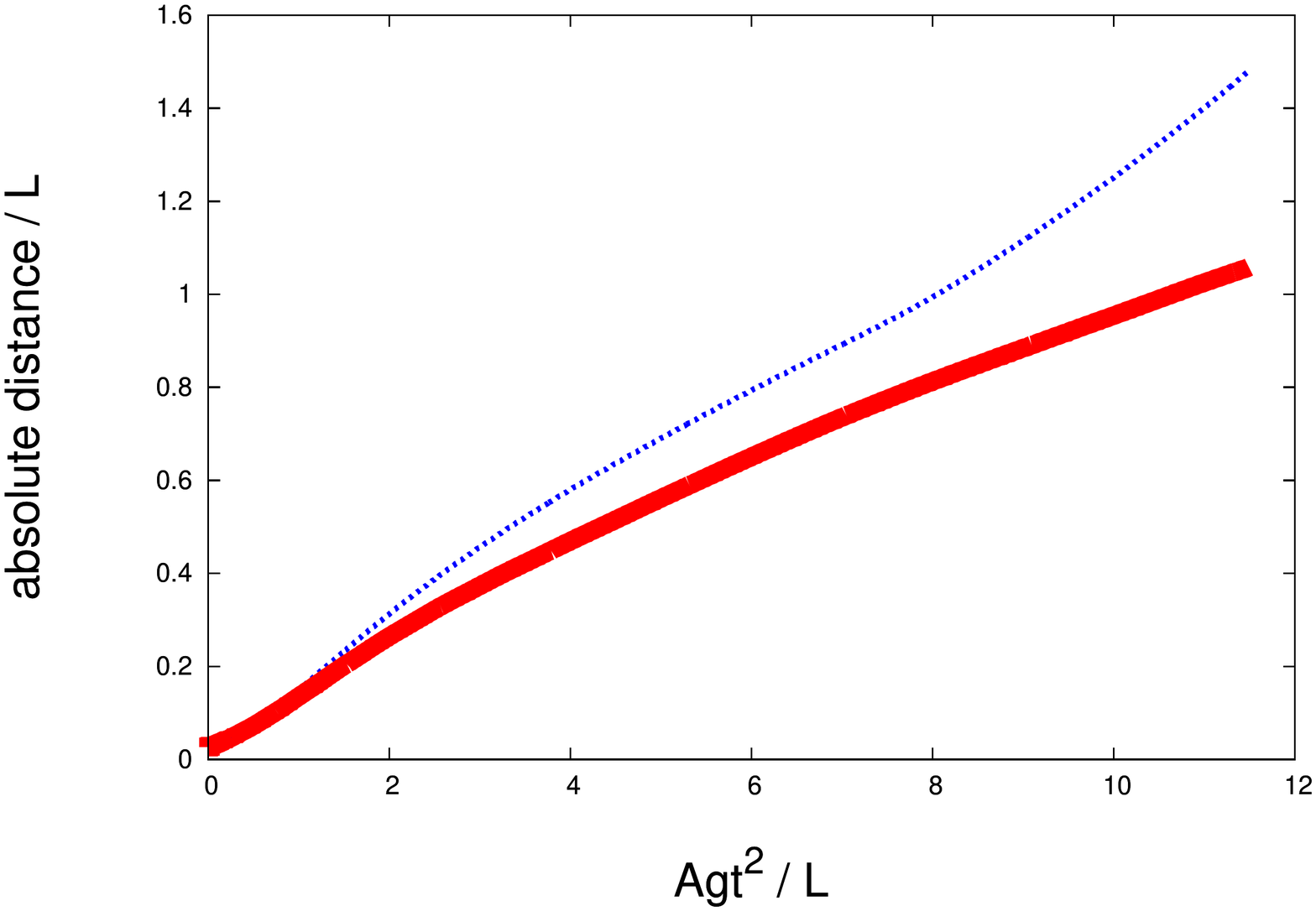}
    \includegraphics[width=0.35\textwidth,angle=0,clip=true]{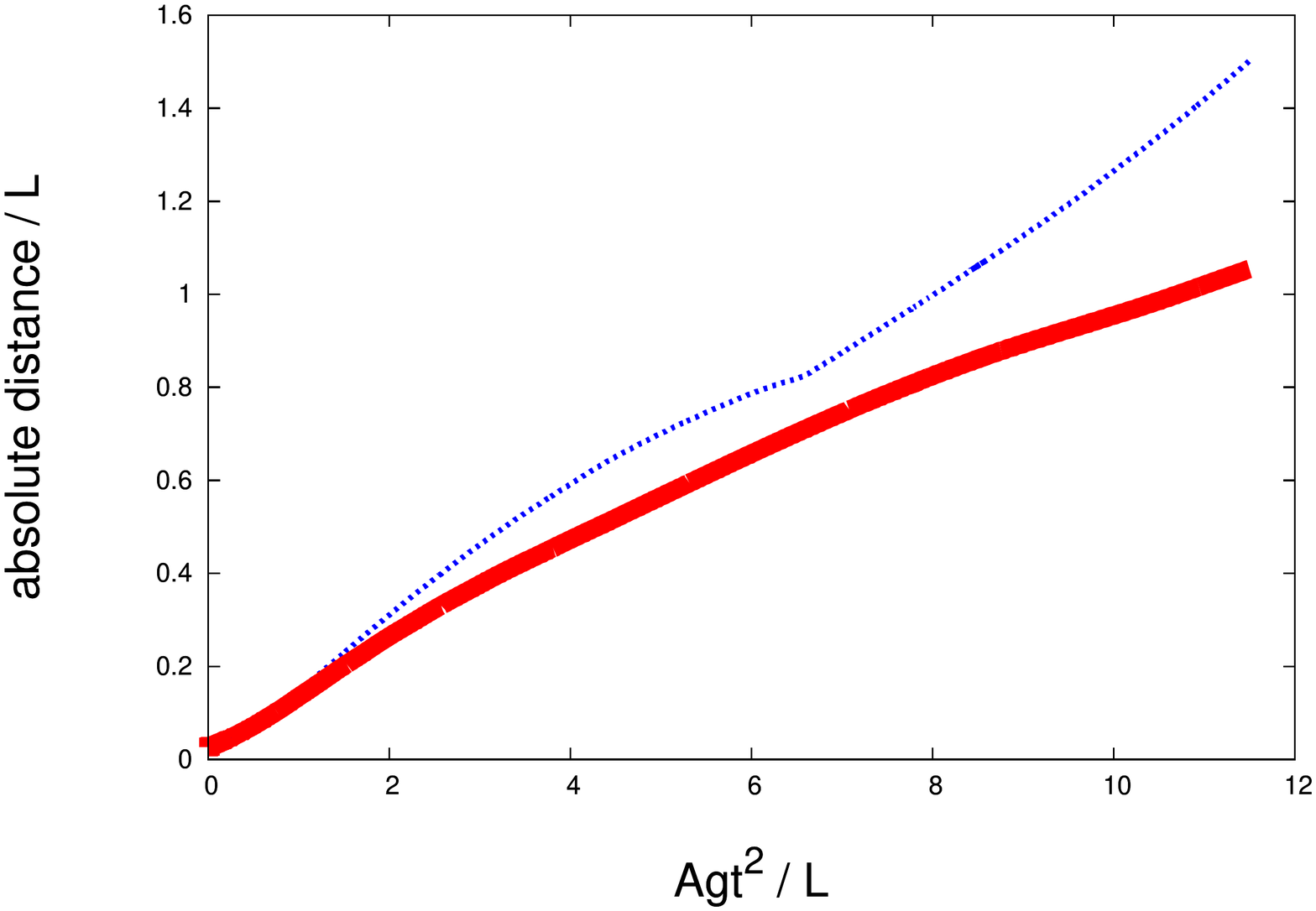}

    \includegraphics[width=0.35\textwidth,angle=0,clip=true]{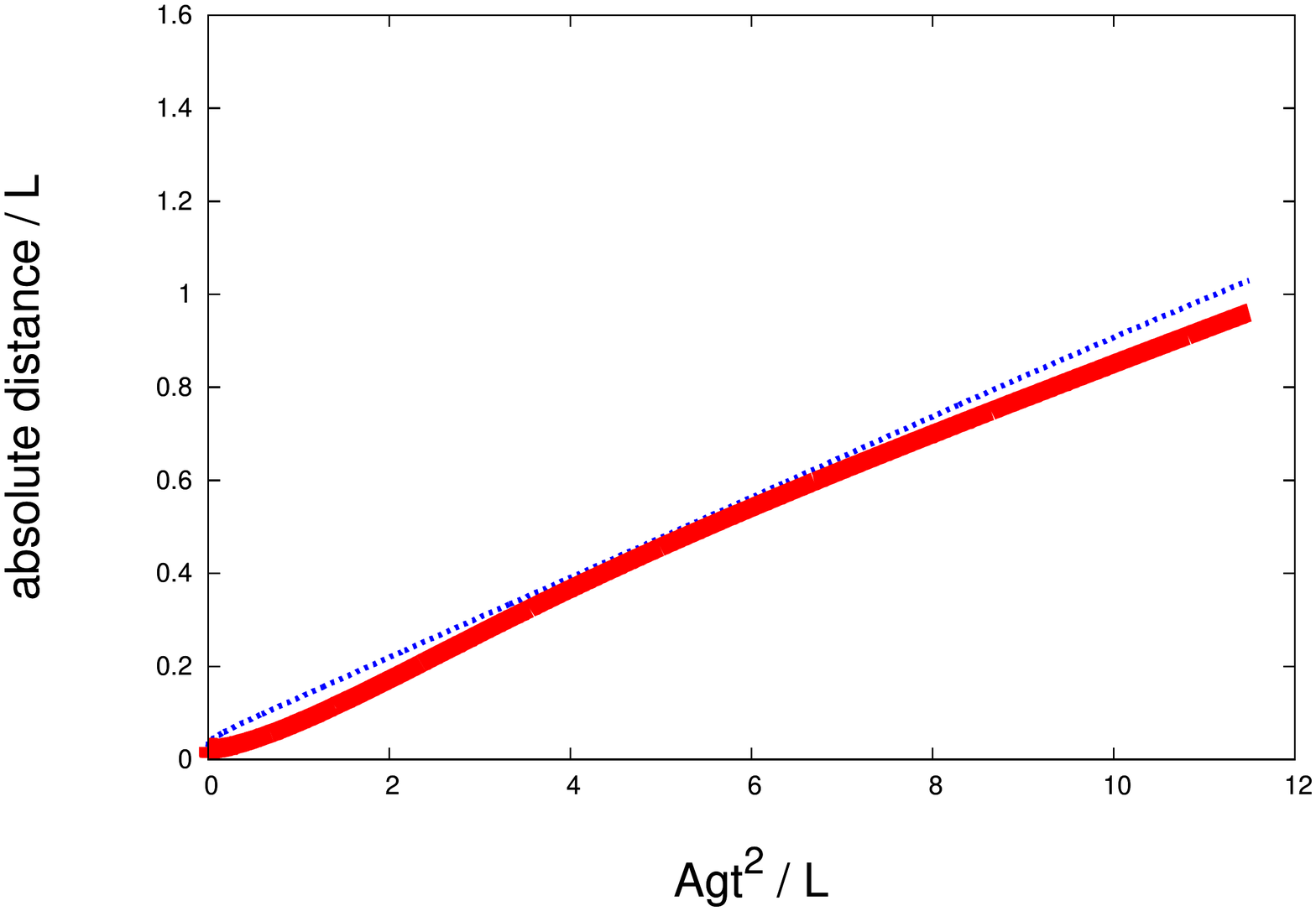}
    \includegraphics[width=0.35\textwidth,angle=0,clip=true]{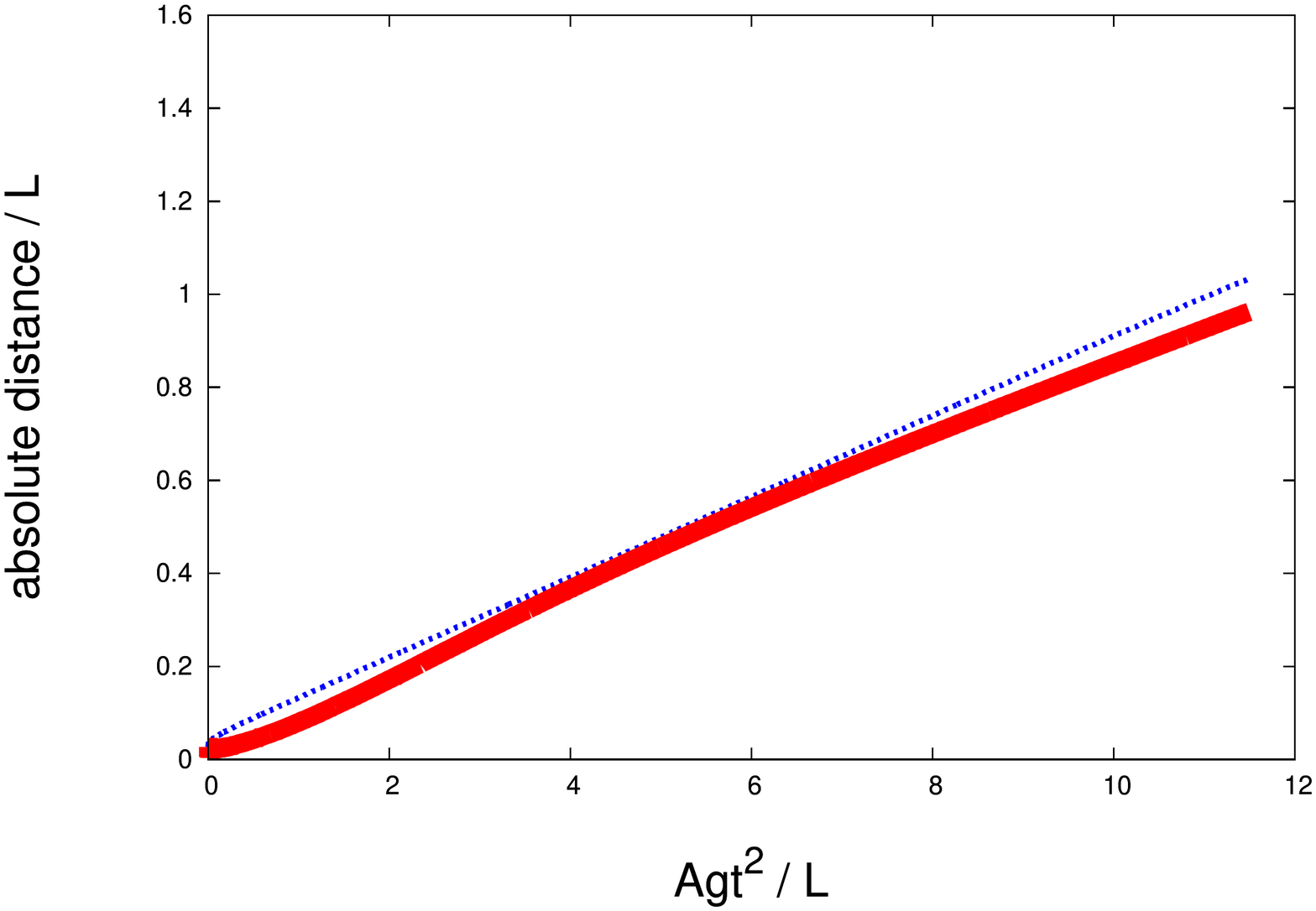}
    \caption{Dependence of positions of tips of RT spikes and bubbles
      on model physics. Absolute distances normalized to the
      perturbation wavelength, $L$, for the RT spikes are shown with
      solid and dashed lines, respectively, as a function of
      normalized time, $A g t^2 / L$. Here $A$ is the Atwood number
      corresponding to the fluid densities outside the smeared
      interface profile. The mesh resolution is L160. (a)
      Hydrodynamics only. (b) Hydrodynamics with self-generation of
      magnetic fields. (c) Hydrodynamics with thermal conduction. (d)
      Hydrodynamics with self-generation of magnetic fields and
      thermal conduction. See Sect.\ \ref{s:lsmixing} for details.}
    \label{f:tips}
  \end{center}
\end{figure*}

\begin{figure*}[htbp!]
  \begin{center}
    \includegraphics[width=0.45\textwidth,angle=0]{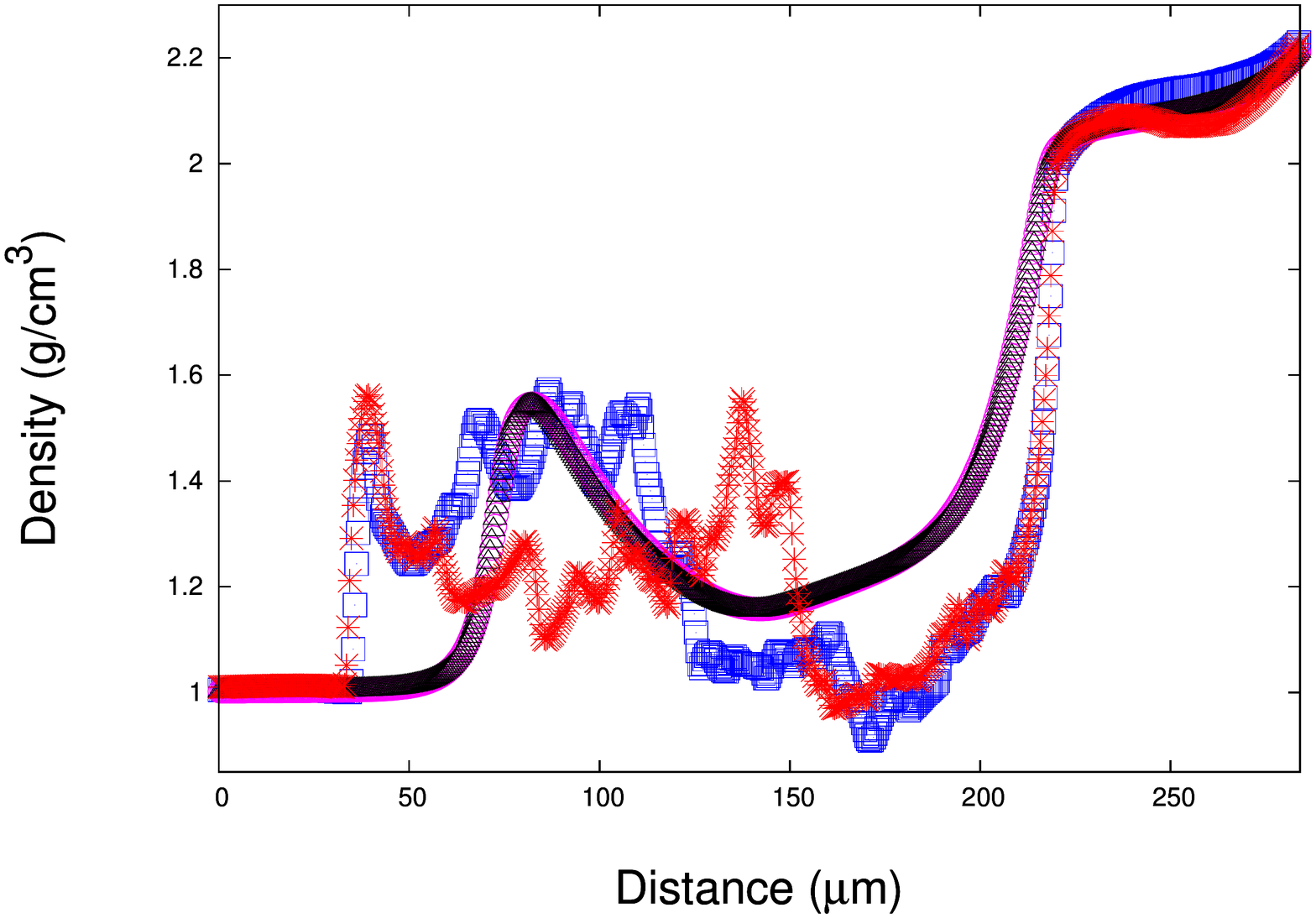}
    \includegraphics[width=0.45\textwidth,angle=0]{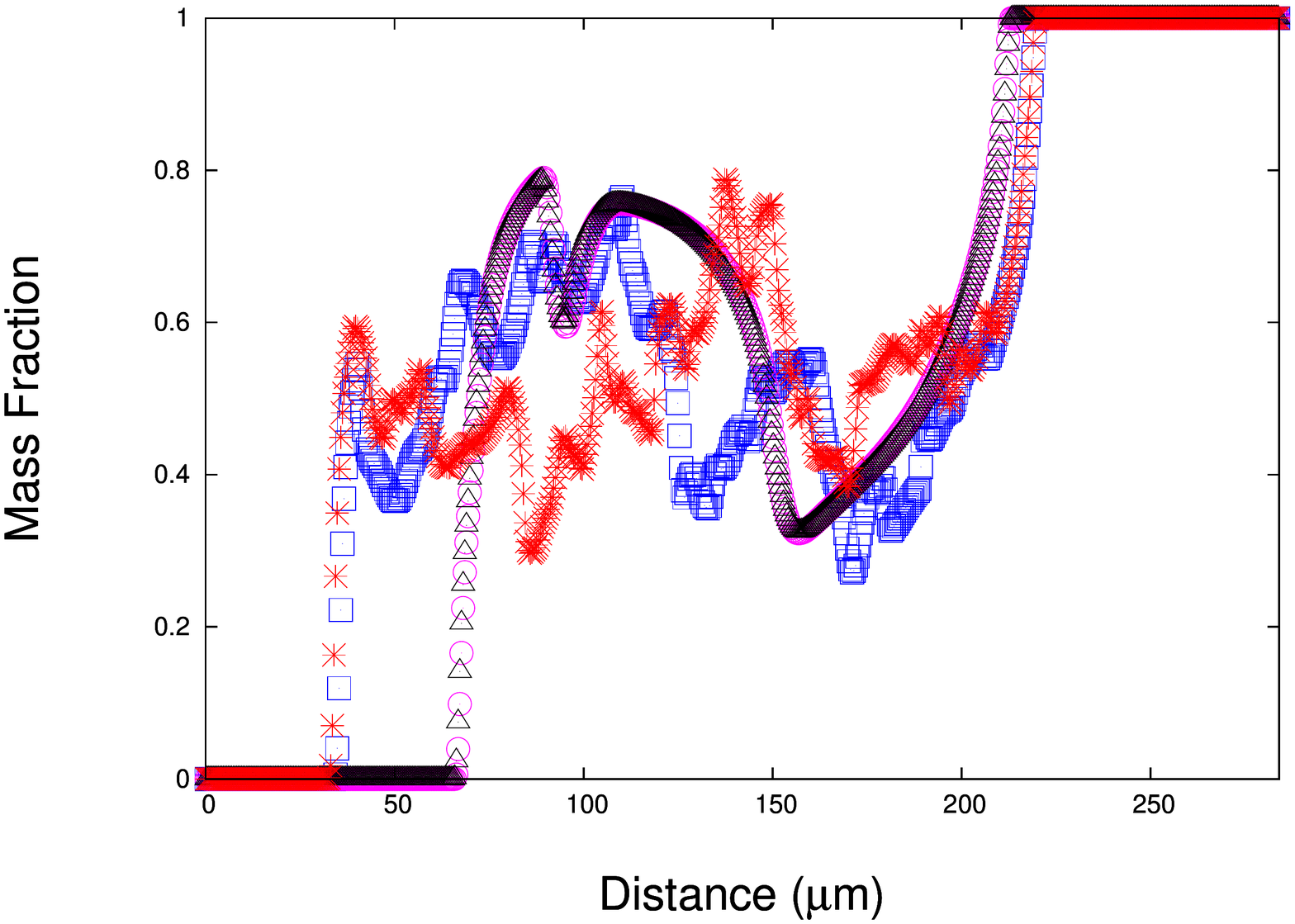}
    \caption{Dependence of laterally averaged structure of the RT
      mixed region on model physics. Lateral averages of density (left
      panel) and heavy fluid tracer mass fraction (right panel) are
      shown as a function of vertical distance at the final time for
      models with hydrodynamics only (squares), hydrodynamics with
      self-generation of magnetic fields (asterisks), hydrodynamics
      with thermal conduction (circles), and hydrodynamics with
      self-generation of magnetic fields and thermal conduction
      (triangles). See Sect.\ \ref{s:lsmixing} for details.}
    \label{f:avps160}
  \end{center}
\end{figure*}

Figure \ref{f:tips}
shows the evolution of positions of tips of RT spikes (where the
lateral average of the tracer reaches $0.01$) and bubbles (where the
lateral average of the tracer decreases to $0.99$) in L160 models with
different physics. By comparing models without (top row of panels in
Fig.\ \ref{f:tips}) and with conduction (bottom row of panels in
Fig.\ \ref{f:tips}), one can clearly see that the mixed layer 
growth is chiefly controlled by thermal conduction. For example, the
mixed layer width reaches about 185 $\mu$m in a pure hydrodynamic model
and is reduced about 20\% (mixed layer width $\approx 146$ $\mu$m) when
conduction is allowed for (Fig.\ \ref{f:avps160}).
The observed reduction of the RTI growth is qualitatively consistent
with theoretical predictions (see \cite{atzeni+04} and references
therein). The effect is largely due to reduced speed of the RT spike;
positions of bubble tips remain essentially unchanged. In the case of
non-conducting models, the mixed layer widths are very similar and
self-magnetization leads to only slight increase of the mixed layer
width (to about 186 $\mu$m).

The relative importance of thermal conduction compared to the
contribution of self-generated magnetic fields is confirmed by close
similarity of lateral averages of the density in thermally conducting
models without and with self-generation of magnetic fields (see
Figs.\ \ref{f:avps}c and \ref{f:avps}d), respectively). When thermal
conduction is inefficient, the role of self-generated magnetic fields
on the overall RTI growth appears negligible with the mixing layer
width essentially identical to that obtained in the pure hydro case
(cf.\ panels (a) and (b) in Fig.\ \ref{f:tips}).

In passing we would like to note that that above conclusions remain
unchanged when analyzing relatively coarse L40 models. We believe that
our best resolved simulations truthfully capture physics effects and
are largely unaffected by numerical diffusion.
\subsubsection{Mixing on small scales}\label{s:ssmixing}
We found that in the case of self-magnetized non-conducting flows much
richer flow structure is created compared to pure hydro situations
(see Figs.\ \ref{f:density35ns}b and \ref{f:density35ns}a,
respectively), presumably due to presence and subsequent decay of
additional higher-order modes. However, this additional small scale
structure is no longer present when conduction is allowed for
(Fig.~\ref{f:density35ns}d). We conclude that thermal conduction
dominates the evolution of RTI on all scales and that self-generated
magnetic fields play a relatively unimportant role in the evolution.
\subsubsection{Characteristics of the self-generated magnetic field}
In simulations that incorporate self-generation of magnetic field
without thermal conduction, the magnetic field at $t=35$ ns is about
${7.5}$ MG. This corresponds to a maximum magnetic pressure on the
order of $1.4$ times the thermal pressure (plasma
$\beta\approx{0.72}$). Furthermore, at intermediate times following
saturation of the field, we found magnetic fields locally reaching
$11$ MG ($\beta\approx{9.1{\times}{10^{-2}}}$). This can be compared to
magnetic field strength of approximately $1.7$ MG ($\beta\approx{49}$)
found when the thermal conduction is allowed for. The magnetic field
is lower in this case because magnetic source terms depend on density
and temperature gradients, which are quickly decreased due to
efficient thermal transport. Also, the field strengthens gradually and
evolves smoothly both in time and space.

Using similar initial conditions but with the pre-existing magnetic
field perpendicular to the interface (aligned with the direction of
gravity), \citet{fryxell+10} has demonstrated that even relatively
weak fields can significantly modify the RTI morphology on large
scales. In particular, they found that the RT mushroom cap does not
form in the case with a pre-existing magnetic field for
$\beta={2500}$. However, the resulting morphology of their model (see
Fig.\ 4 in \cite{fryxell+10}) does not resemble complex structure of
our non-conducting model with self-generated field
(cf.\ Fig.\ \ref{f:density35ns}b). Instead, it is similar to
morphology of featureless RT spikes seen in our thermally conducting
models at early times (cf.\ Fig.\ \ref{f:density20ns}c). We conclude
that pre-existing smoothly ordered fields produce qualitatively
different effects than self-generated magnetic fields and the results
of such studies should not be in general used to interpret the
evolution of self-magnetizing systems with zero-field initial
conditions.
\subsubsection{Anisotropy of heat fluxes}
The degree of anisotropy of thermal transport due to magnetic field
can be estimated based on the value of parameter $\chi$
(cf.\ Eqs. \ref{e:kappa_e} and \ref{e:kappa_i}), which depends on the
magnetic field strength. The heat transport is essentially isotropic
in situations with $\chi \ll 1$, it becomes progressively more
anisotropic as the field strengthens, and one may expect new effects
due to anisotropy as $\chi$ reaches values of order 1 
\cite{braginskii65,nishiguchi02}. In our subset of models without
conduction (that produces the strongest fields), we find $\chi\approx
0.001$, which indicates that heat fluxes are essentially isotropic and
effects due to anisotropic thermal conduction are negligible.
\subsubsection{Modified RTI growth}\label{s:additionalmode}
As we mentioned in Sect.\ \ref{s:hydmag}, the important effect of the
magnetic field generated in the absence of thermal conduction is the
creation of an additional RTI mode. Presence of this mode is clearly
visible by comparing the results of a purely hydrodynamic model at
$t=20$ ns (Fig.\ \ref{f:density20ns}a) to the corresponding
non-conducting model with self-generated magnetic field
(Fig.\ \ref{f:density20ns}b). \citet{nishiguchi02} also found an
additional mode in his simulations of RTI in the ICF setting. However,
contrary to Nishiguchi's interpretation, thermal conduction is
unnecessary for creation of additional modes in our model.
Nishiguchi's mechanism requires strong magnetic fields in order to
make the conduction anisotropic and its perpendicular components
suppressed.

The essence of Nishiguchi's mechanism is that it is the suppressed
conductivity that causes the laser heat flux decrease at the
corresponding section of the interface. In consequence, the ablation
rate is locally decreased and hence the acceleration that drives
RTI. In effect, the RTI growth near the spots with maximal $B$ is
decreased, and so, similar to our model, an additional mode, not
present in the pure hydro RTI, is generated. As we pointed in
Sect.\ \ref{s:hydmag}, the self-generated magnetic field is most
likely directly responsible for creation of additional perturbation
modes in our setup. In this interpretation, thermal conduction may
only enhance the effect; our dynamical mechanism must be present in
the Nishiguchi's model as well. Unfortunately, Nishiguchi does not
discuss the effects of self-generated magnetic fields in the absence
of thermal conduction, and the data presented in his paper is
insufficient to judge which of the two mechanisms dominates in his
model. The problem of generation of additional RTI modes in the ICF
regime deserves a careful study of individual physics effects, similar
to presented in this work.

The condition for either of the two mechanisms of the additional mode
generation to be effective is that $\beta$ should become $\lesssim 1$.
Suitable physical conditions for this are high enough temperatures (a few
hundred eV) so that resistive diffusion is small enough to allow the
magnetic field to grow to MG-level values while localized close to the
interface region (we estimate the resistive diffusion below, see Sect.\
\ref{s:resistive}); and substantial acceleration to keep the interface
between the light and the dense plasmas thin. Both of these conditions
appear satisfied in Nishiguchi's setup.
\subsection{Technical aspects}
\subsubsection{Numerical mesh convergence}
We found that in the case of pure hydrodynamics the large scale flow
features are well-resolved already in a relatively coarse L40 (40
zones per the perturbation wavelength) model.  As we increase the
resolution, the shear flow near the interface between the spike
material and the ambient medium becomes better resolved allowing for
more vigorous development of structure on small scales.

The increase of the amount of structure with the resolution and the
corresponding lack of convergence on small scales is a well-known
characteristic of RTI models. This is chiefly due to a lack of physics
(such as viscosity) that would control flow development on small
scales but above the mesh resolution limit. In this situation, one has
to resort to a convergence metric insensitive to variations on small
scales such as lateral averages.

The lateral averages of density for simulations of pure hydro and
hydro with self-generated magnetic field simulations in
Figs.\ \ref{f:avps}a and \ref{f:avps}b show that the spike position
has converged by the L80 resolution. Oscillatory changes of the
lateral averages seen near the spike in the L160 model with
self-generated magnetic fields indicate magnetic fields are
responsible for substantially different density structure in this
model compared to pure hydro. The convergence in this model also
appears the slowest. By contrast, the lateral averages for the two
simulations that incorporate thermal conduction are nearly identical
at all resolutions and appear to have converged already in the L40
model.

The flow structure on small scales is significantly richer in the case
magnetic fields are self-generated and thermal conduction is absent
(see Fig.\ \ref{f:density35ns}b). This leads to visibly slower
convergence of lateral averages of density with mesh resolution
(Fig.\ \ref{f:avps}b). The enhanced small scale structure is likely
due to perturbations induced by the additional RTI mode, and it is
conceivable that similar phenomena develop on smaller scales leading
to perturbations at higher frequencies as the interface becomes more
convoluted. We defer analysis of this interesting possibility to the
future work.

Finally, the overall symmetry of the flow is well-preserved on large
scales at all times, although the morphology appears asymmetric on
small scales (Fig.~\ref{f:density35ns}b). Such asymmetries are
typically observed in RTI simulations in the nonlinear phase when
otherwise negligible numerical discretization errors become
sufficiently large to perturb the flow.
\subsubsection{Heat flux limiting}\label{s:heatlimit}
When electron heat flux according to Braginskii expressions becomes
extreme, reaching a few percent of the free-streaming value,
Braginskii's value is no longer justified and flux-limiters are used:
\[
| \vec{q}_T^{\,e} | = \mathrm{min}\left( |{\mbox{\boldmath$\kappa$}}_e\cdotp\nabla T_e|, f n_e T_e
(T_e/m_e)^{3/2}\right) ,
\]
with $f$ usually chosen $f\in[0.03; 0.1]$ \citep{malone+75}.  At still
larger gradients (of the temperature in part), when thermodynamical
quantities change significantly over distances of order of electron
mean free path $l_e=\tau_e (T_e/m_e)^{1/2}$ MHD description becomes
invalid as well.  In our simulations the temperature along the
interface varied in $[4{\times}10^5; 9{\times}10^5]$K range (about
$[3.8{\times}10^4; 1.3{\times}10^6]$K throughout the domain). For a
representative value of the density of $1.5$ g cm$^{-3}$
($n_e=4.5{\times}10^{23}\mathrm{cm}^{-3}$), we have
$l_e\in[3.25;13.4]{\times}10^{-9}\mathrm{cm}$, which is by far shorter
than the interface thickness in our simulations. Thus MHD description
of the problem is well justified.  The ratio of the Braginskii's value
of the electron heat flux to the free-streaming value at $|\nabla
T_e|=10^9$ K cm$^{-1}$ (about the largest value we had in the
simulations) and $\kappa_e=2.8{\times}10^{9}$ g cm s$^{-3}$ K$^{-1}$
(at $T_e=9{\times}10^5$ K) is at most $\approx
1.3{\times}10^{-4}$. Consequently there was no need to use flux
limiters in our simulations.
\subsubsection{Computational limitations due to thermal transport}
We have also found that conducting models become prohibitively
expensive at resolutions higher than L160 due to a time step limited
by the thermal conduction timescale,
\[
\Delta t \leq \frac{1}{2} \frac{{\Delta x}^2}{ D},
\]
where $D$ is the temperature diffusion coefficient
(cf.\ Eq.\ \ref{e:diffcoeff}). For the current application, the
thermal diffusion is dominated by free electrons and shows strong
dependence on temperature, $D \propto T^{5/2}$. This imposes a
practical limit on maximum temperatures attainable in our
computational models using explicit time integration to
$1{\times}10^6$ K. Modeling situations at higher temperatures will
require using implicit solvers for thermal conduction for the
simulations to remain computationally feasible and numerically
accurate.
\subsubsection{Magnetic field growth limiter}
Under high-energy density conditions considered here, magnetic fields
are generated very quickly. In the absence of thermal conduction and with
only minimal numerical diffusion due to zero initial velocities,
temperature and density gradients remain large and close to their
initial values for prolonged amounts of time. At the same time, the
generated field is neither distributed (a numerical Z-pinch-like
effect occurs due to assumed 2-D geometry) nor advected. Combination
of those two effects occasionally results in production of magnetic
fields so strong that the resulting magnetic pressure leads to a
complete evacuation of computational zones and unphysical
solutions. In order to avoid such situations, we limit the magnetic
field growth to a small (typically a few per cent) fraction of the
existing field, provided the field is sufficiently strong,
\[
\Delta\vec{B} =\left\{\begin{array}{ll}
\min\left[ C_\mathrm{Bmax} \vec{B} , \Delta t \left(\frac{\partial\vec{B}}{\partial t}\right)
  \right] & \mbox{ if ($P_\mathrm{mag} \geq C_{\beta} P_\mathrm{therm}$),} \\
         \Delta t \left(\frac{\partial\vec{B}}{\partial t}\right) & \mbox{ otherwise.}
           \end{array}\right.
\]
where we have set $C_{\beta}=1{\times}10^{-3}$. We experimented using
various values of the field growth limiter, $C_\mathrm{Bmax}$, and
found that $C_\mathrm{Bmax}=3{\times}10^{-3}$ was sufficiently small
to produce stable solutions in all cases with exception of L160
non-conducting model, which required
$C_\mathrm{Bmax}=1{\times}10^{-3}$.

We believe the above field growth limiting strategy is justified as
the procedure is only applied locally and does not affect large scale
features of the solution. Also, we anticipate the above problem will
not be present in 3-D due to qualitatively different geometry of the
generated field.
\subsection{Resistive effects}\label{s:resistive}
One can provide a rough estimate of magnetic field resistive diffusion
by considering a one-dimensional diffusion problem with magnetic field
source term of the form $f(t)g(x)$, located in a vicinity of the
interface at $x=0$. We take $g(x)=(\pi b^2)^{-1/2} \exp(-x^2/b^2)$,
where $b$ is the characteristic thickness of magnetic field generating region
($b\approx3$ $\mu$m in our L160 model with no heat conduction,
and $b\approx 15-20$ $\mu$m in the models with Braginskii heat
conduction.)

The solution of the diffusion problem with constant isotropic magnetic
diffusivity $D_m$ and the above source term
\[
\partial B(x,t)/\partial t= \nabla(D_m\nabla B)+f(t)g(x),
\]
and with $B(x,0)=0$ is\
\[
B(x,t)=\int_0^t d\tau f(t-\tau) \frac{\exp(-x^2/(b^2+4D_m\tau))}{[\pi
(b^2+4 D_m\tau)]^{1/2}}.
\]
When $D_m\to 0$, $B(x,t)\propto\int_0^t f(\tau)\, d\tau$, is similar at
all $x$. We thus take $f(t)=f_0=const$ to reproduce approximately
linear in time growth of $B(x,t)$ at the interface at early times. At
late times, $t\!\gg\! b/D_m$, the distribution of magnetic field has
characteristic thickness of $(b^2+4D_mt)^{1/2}$, whereas
$B(0,t)/B(0,t)|_{D_m=0}\approx b/\sqrt{D_mt}$.

Magnetic diffusivity coefficient $D_m$ can be expressed in terms of
Braginskii's $\alpha$ coefficients as $D_m=\frac\alpha{4\pi}\left(
\frac c{en_e}\right) ^2$. As with heat conductivity, resistivity (and
hence magnetic field diffusion) is essentially isotropic in our
problem since $\chi=\omega_e\tau_e\ll 1$.  In our simulations, $D_m$
varies approximately between $1.7{\times}10^4$ and $0.6{\times}10^4$
cm$^2$ s$^{-1}$ along the interface. The corresponding temperatures
are approximately $4{\times}10^5$ K and $9{\times}10^5$ K,
respectively, with the electron number density, $n_e\approx
4.5{\times}10^{23}$ cm$^{-3}$. For the estimated average value of the
diffusivity coefficient, $D_m=1{\times}10^4$ cm$^2$ s$^{-1}$ (or
$D_m=1{\times}10^5$ $\mu$m$^2$ ns$^{-1}$), it takes $10^{-3}$ ns for
the field to diffuse across the distance of 10 $\mu$m. Consequently,
for the present application, we can expect resistivity starting to
affect the magnetic field evolution for time intervals longer than
about $10^{-3}$ ns.

At $t=20$ ns and assuming characteristic thickness of the interface in
our simulations with Braginskii heat conduction, $b=15$ $\mu$m, we
estimate the field will be reduced by a factor of
$\approx\sqrt{D_mt}/b \approx 100$ compared to the non-resistive
problem we consider in this work. One should note that at higher
temperatures which are achieved in experiments \cite{kuranz+10}, the
role of resistivity will be smaller.

Another effect related to resistivity that we omitted in the present
work is the thermoelectric heat flux, $\vec{q}^{\,e}_u=-\frac T e\frac
c{4\pi} \beta_0\mathrm{rot}\vec{B}$ \cite{braginskii65}. Adopting the
values for the maximal magnetic field strength and the characteristic
length scale of the region where this field is localized at $t=20$ ns,
and depending on which physics is included, we estimate
$|\vec{q}^{\,e}_u|\in[10^{13};10^{17}]$ erg cm$^{-2}$ s$^{-1}$. For
instance, if both resistivity and heat conduction are included,
characteristic $B$ is about $5\times 10^3\:$G, spread over 3 mm,
leading to $|\mathrm{rot}\, \vec{B}^{}|\approx 2\times
10^4$ G cm$^{-1}$ and the lower bound for $|\vec{q}^{\,e}_u|$
in the estimate above; whereas not taking resistivity into account
would yield $|\mathrm{rot}\, \vec{B}_{}|\approx 2\times
10^8$ G cm$^{-1}$ and the upper bound for the
$|\vec{q}^{\,e}_u|$ given.  On the other hand, using our estimates
presented in Sect.\ \ref{s:heatlimit} above yields thermal heat flux
$|\vec{q}^{\,e}_T|\approx 7{\times}10^{17}$ erg cm$^{-2}$ s$^{-1}$.

We conclude that taking the thermoelectric heat flux into account
should not change our simulation results significantly. Also, since
magnetic fields generated in the thermally conducting model are
already relatively weak, the resistive effects should not change RTI
evolution in any qualitatively new way. These expectations will be
tested against RTI models with resistivity taken into account in the
next paper in the series.
\subsection{Comparison with previously published experimental and theoretical work}
\paragraph{Experiments}
In a series of HED experiments on the OMEGA laser, \citet{manuel+12}
studied the generation of magnetic fields induced by the
Rayleigh-Taylor instability. Using the proton radiography technique
and path integrated measurements they were able to estimate the
magnetic field strength. Their estimates of the \emph{maximum} field
strength were obtained by assuming the field was localized in a region
with a thickness on the order of the perturbation amplitude. The
perturbation amplitude grew exponentially in time at a rate about 2.2
ns$^{-1}$, and the magnetic field reached its maximum strength of 0.15
MG at $t=1.5$ (the end of the linear RTI stage). The authors did not
provide estimates for the field strength at later times.

Megagauss-level magnetic fields were observed for the first time in
HED experiments by \citet{gao+12}. The target used was a 15 $\mu$m
thick polystyrene foil, compared to 21 $\mu$m thick foil used in
\citet{manuel+12}. \citet{gao+12} were able to measure the fields well
into nonlinear RTI regime, up to 2.56 ns after the laser beam arrival
at the target. At that time the foil was broken apart by the
instability. Maximal field strengths up to 2 MG were inferred from
path-integrated measurements. \citet{gao+12} also considered a 25
$\mu$m thick foil and found sub-MG fields; the foil stayed unbroken,
suggesting that the fastest field growth was attained in the late,
highly nonlinear stage of RTI.

The relevant model for comparison to these experiments is our
conducting model with self-generated magnetic field (see
Sect.\ \ref{s:hydcondmag}). In this model, the \emph{maximum} magnetic
field strength initially rapidly increases to $83$ kG at $t=1$ ns,
after which the growth slows and fields on the order of $103$ kG are
obtained at $t=2$ ns. Between $t=2$ ns and $t=10$ ns the average rate
of field growth is $27$ kG ns$^{-1}$. We find these results to be in
good agreement with the experimental works, considering the
differences between the initial conditions in our model (motivated by
the Kuranz \etal experiment) and the actual conditions of the
experiments described above. The field growth rate is higher in these
experiments due to a target made of a thin foil. This is because for
the same laser drive parameters, one obtains greater accelerations and
so also greater RTI growth rates for lighter targets. (We estimate the
characteristic RTI growth rate in our conducting model with
self-magnetization to be about 0.14 ns$^{-1}$ compared to 2.2 ns$^{-1}$
obtained in the
Manuel \etal experiment.) Indeed, \citet{gao+12} reported that the
foil remnants had a velocity of $(3\pm 1)\times 10^7$ cm s$^{-1}$ at
$t=2.56$ ns, implying average acceleration of $\approx 10^{16}$ cm
s$^{-2}$. This acceleration is nearly 2 orders of magnitude larger
than the acceleration adopted in our study ($g=2\times 10^{14}$ cm
s$^{-2}$), which is representative of the Kuranz \etal experiment. In
what follows, we discuss additional differences between our setup and
the experimental designs.

No seed perturbations were machined on the foil used in the Gao \etal
experiment. The authors reported $\approx 82$ $\mu$m wavelength
perturbations dominating at $t=2.11$ ns, and $\approx 115$
$\mu$m perturbations at $t=2.56$ ns. These are comparable to
$\lambda=71$ $\mu$m perturbations used in \citet{kuranz+10} experiment
and in this study. Similar perturbation wavelengths were used in
\cite{manuel+12}, with $\lambda=120$ $\mu$m (and with an amplitude of
$a_0=0.27$ $\mu$m). The differences in wavelengths between various
setups are relatively small and unlikely to cause significant
deviations in magnetic field growth rates. Reported densities of the
target material during the experiments/simulations are also very
close. The main factors that we link to the difference in $B(t)$ are
the interface acceleration, the temperatures near the interface (these
strongly influence transport properties, and enter the Biermann
battery term directly as the $\nabla T_e$ factor), differences in the
flow structure in ablative RTI setup compared to our classical RTI
configuration
.

\paragraph{Theoretical and computational studies}
We begin a comparison of our simulation results to those reported by
other groups with one of the early numerical studies of magnetic field
generation by \citet{mima+78}. The hydrodynamical process studied by
that group was not RTI in its classical meaning, as no gravity or
sufficiently smooth (on interface thickness scale) global acceleration was
present. Instead, the authors studied equilibration in two setups. The
first configuration included discontinuities in both pressure and
temperature (\lq\lq implosion setup\rq\rq ). In the second
configuration the plasma was initially in pressure equilibrium but
with a temperature discontinuity. The process of hydrodynamical
equilibration was accompanied by the perturbations growing on
initially flat interface separating two sections of the computational
domain that contained plasmas at constant but different pressures and
temperatures. The temperature varied between 50 eV and 200 eV, with
the maximum temperature higher by a factor roughly 2.5 than the
maximum temperature at the interface in our model ($\approx 80$ eV). 
The magnetic field
growth was computed only using the Biermann battery, and the maximal
fields obtained in the above two configurations were 3 MG and 0.3 MG,
respectively.  The magnetic field saturated after one half the sound
crossing time of the domain, and thermodynamic equilibrium was
achieved by about that time in both setups. Differences in the flow
morphology obtained by \cite{mima+78} and reported here are
significant and do not allow for meaningful comparison of the results
of the two studies.

A setup similar to ours was studied more recently by
\citet{nishiguchi02}. The simulation code used by Nishiguchi solved
the full set of Braginskii's equations in 2-D, and modeled laser
energy deposition in the 50 $\mu$m thick target of
CH$_2$ composition. The target surface was perturbed using a single mode
sinusoidal perturbation with $\lambda=100$ $\mu$m.  Results of two
types of simulations were reported. In the first case, transport
coefficients were fixed at their $B=0$ values; in the second case,
these coefficients depended on the magnetic field according to
Braginskii's formulation. Maximal B-field values reported for the
above two cases at $t=3.4$ ns were 2 MG and 8 MG, respectively. Such a
substantial difference in field strengths was due to the increase of
the temperature gradient in the case where thermal conductivity
depended on the magnetic field (i.e., heat flux was reduced in the
direction orthogonal to the magnetic field). This makes the
Nishiguchi's system substantially different from ours, because in our
model the magnetic field was relatively weak and thermal conduction
remained essentially isotropic.

The discrepancy between our models and Nishiguchi's simulations stems
from differences in temperature at the interface region. Although
Nishiguchi does not provide the temperature near the interface in his
model, he reports the maximal temperature in his simulations as
$2.6\times 10^7$ K, which is nearly 20 times higher than the maximum
temperature recorded in our models. This results in
$\chi=\omega_e\tau_e$ in Nishiguchi's simulations being a factor of
200 greater than ours for the same magnetic field strength ($\tau_e$
scales approximately as $T_e^{3/2}/\bar{Z}$), with $\chi\approx 0.5$
for $B=2$ MG. The thermal conductivity is thus reduced as
$\kappa_\perp/\kappa_{||}\approx 0.55$. In consequence, the
temperature diffusion across the field lines is suppressed resulting
in thinner interface and faster magnetic field growth. As the field 
progressively strengthens, thermal transport is reduced even more, 
increasing the field growth rate still further.

The higher temperatures found in the Nishiguchi's study appear to be
caused by both target design and driving energy. The target proposed
by Nishiguchi is thinner and is continuously driven by a much stronger
laser source. The laser intensity used by Nishiguchi was a constant
$2\times10^{15}\:\mathrm{W\, cm^{-2}}$, while the laser drives used in
experiments were weaker and lasted only for a short amount
of time ($9\times 10^{14}\:\mathrm{W\,cm^{-2}}$ for 1 ns in
\cite{kuranz+10}; $4\times 10^{14}$ for 2 ns in
\cite{manuel+12,gao+12}). Although \cite{nishiguchi02} does not
provide the effective acceleration value, assuming a 2 MG magnetic
field and a few micron thick interface, we can use the results shown
in his Fig.\ 1 to estimate $g\approx 10^{16}\:\mathrm{cm\,
  s^{-2}}$. This estimate is consistent with the results presented by
\cite{gao+12}, and again much higher than in our model.

Apart from the computer simulations discussed above,
\citet{nishiguchi02} and \citet{manuel+12} independently provided
estimates for the resistive diffusion. In particular,
\citet{manuel+12} estimated the resistive diffusion time on the order
of 1 ns. This implied a reduction of the magnetic field by a factor
2.5 compared to the non-resistive result. We note that we used a
different approach to estimate resistive diffusion (see
Sect.\ \ref{s:resistive}). Specifically, we did not substitute the
spatial differential operator with a wavenumber as Nishiguchi and
Manuel \etal did. This difference in the analysis method is especially
important in case of thin interfaces and at early times.  This is
because, and as we have shown, during the initial stages magnetic
field mostly diffuses away from the interface, not along it. This
means that the magnetic field is distributed in space with
characteristic thickness different than that of the interface or the
interface perturbation wavelength, invalidating the approach adopted
by Manuel \etal. In consequence, Manuel \etal overestimated the
diffusion timescale and underestimated the field reduction factor.

We conclude that the magnetic field strength and its growth rate
obtained in our numerical simulations are consistent with those
obtained in simulations by other groups and in the recent HED
experiments provided differences in the problem setting are taken into
account.
\section{Conclusions}\label{s:conclusion}
Motivated by discrepancies between the results of the Rayleigh-Taylor
instability (RTI) high-energy density laser experiments and computer
simulations, we investigated the effects of physics previously not
included in computer models on the instability growth and morphology
of the flow.

Our physics model is based on the Braginskii formulation
\cite{braginskii65}, but we do not account for resistive effects.  We
implemented anisotropic thermal conduction and source terms for
self-generation of magnetic fields in the \FLASH-based
\Proteus\ code. The new computational modules were subjected to a
verification process and produced correct results for several test
problems.

Using the verified code, we performed a series of simulations of a
single mode RTI in 2-D planar geometry for conditions relevant to the
experiments. Models differed in mesh resolution and physics. We found
that in models with thermal conduction the mixed layer width converged
in simulations with 40 mesh cells per perturbation wavelength; twice
higher resolution was required to obtain convergence when the thermal
conduction was absent.

For relatively moderate HED plasma conditions considered in the
present work, we found that thermal conduction inhibits development of
small scale structure and adversely affects the RTI growth due to
strong mass diffusion stemming from heat transport and the related
density gradient reduction. Our result provides a plausible
explanation for the relatively featureless RT spikes characteristic of
the laboratory experiments. However, we do not observe the mass
extensions seen in those experiments, even in simulations with no
thermal conduction that produce relatively strongest RTI growth.

We found that magnetic fields grow up to approximately $11$ MG (plasma
$\beta\approx{9.1\times{10^{-2}}}$) in the absence of thermal
conduction, with the average fields on the order of $2.5$ MG. These
fields do not affect the integral RTI growth rate but do change RTI
morphology on scales comparable to the perturbation wavelength.
Specifically, we found evidence of additional higher order mode
development due to dynamical effect of these fields. Generation of
such a mode was reported by \citet{nishiguchi02} who interpreted the
additional mode as being due to the (perpendicular) thermal conduction
components getting reduced by the field. Importantly, in the
Nishiguchi's model, this field is maximal in the same region of the
interface as in our simulations.

We also found that self-magnetization enhances RTI mixing
on small scales, possibly due to contribution and decay of additional
higher-order modes. Furthermore, we found that magnetic fields are
responsible for deceleration of the flow near and along the spike's
center line resulting in significant ''denting'' of the spike's
surface.

We identified generation of additional RTI mode(s) and changes of the
spike morphology as unique features of our non-conducting
self-magnetized models. This finding suggests that the pre-existing,
smoothly ordered fields produce qualitatively different effects than
those in which magnetic fields are self-generated. We conclude that
the results of studies using such pre-existing fields in general are
not helpful in understanding the evolution of self-magnetizing systems
with zero-field initial conditions.

In our fully integrated model with self-generation of magnetic fields
and thermal conduction, temperature gradient and density gradient were
quickly reduced resulting in weaker magnetic fields, $B\approx{1.7}$
MG (plasma $\beta\approx{49}$). These magnetic fields were too weak to
make heat transport strongly anisotropic or produce additional RTI
modes. More importantly, we found these fields of comparable strength
to self-generated magnetic fields observed recently in HED laboratory
experiments by \citet{manuel+12}.

The present study offers several directions for future research. For
example, 3-D simulations are necessary to describe the dynamics
involved in RTI more truthfully, especially with regard to the process
of competition between hydrodynamic drag and buoyancy
\cite{kane+00,drake12}. The topology of self-generated magnetic fields
in 3-D will also be different and may result in new phenomena, and
their morphology might be modified on small scales due to resistive
effects we did not include in this work. Finally, future studies
should be conducted in the regime more closely matching conditions of
the past and planned high-energy density RTI laboratory experiments.
\section{Acknowledgments}\label{s:ack}
We thank the reviewer for helpful comments on the initial version of
the manuscript which helped improving presentation of our results. FM
and TP were supported in part by the DOE grant DE-FG52-09NA29548 and
the NSF grant AST-1109113. This research used resources of the
National Energy Research Scientific Computing Center, which is
supported by the Office of Science of the U.S.\ Department of Energy
under Contract No.\ DE-AC02-05CH11231. The software used in this work
was in part developed by the DOE Flash Center at the University of
Chicago.
%
%
\bibliographystyle{model1a-num-names}
\bibliography{manuscript}
%
%
\end{document}